\documentclass[12pt]{article}
\usepackage{graphicx}
\usepackage{latexsym,amsmath,amsfonts,amssymb,bm,bbm}
\usepackage{jheppub}
\usepackage[usenames, dvipsnames, svgnames]{xcolor}
\usepackage{tikz}
\usetikzlibrary{intersections, calc, patterns}

\def\CA{{\cal A}}

\def\BE{\mathbb{E}}
\def\BF{\mathbb{F}}
\def\BK{\mathbb{K}}

\makeatother

\begin{document}

%
\preprint{IPMU-14-0363, UT-14-48}
\title{Entanglement Entropy of Annulus in Three Dimensions}

\author[a,b]{Yuki Nakaguchi}
\author[b]{and Tatsuma Nishioka}

\affiliation[a]{Kavli Institute for the Physics and Mathematics of the Universe (WPI)\\
The University of Tokyo,\\
5-1-5 Kashiwa-no-Ha, Kashiwa City, Chiba 277-8568, Japan}
\affiliation[b]{Department of Physics, Faculty of Science\\
The University of Tokyo,\\
Bunkyo-ku, Tokyo 113-0033, Japan}

\emailAdd{yuuki.nakaguchi@ipmu.jp}
\emailAdd{nishioka@hep-th.phys.s.u-tokyo.ac.jp}

\abstract{
The entanglement entropy of an annulus is examined in a three-dimensional system with or without a gap.
For a free massive scalar field theory, we numerically calculate the mutual information across an annulus.
We also study the holographic mutual information in the CGLP background describing a gapped field theory.
We discover four types of solutions as the minimal surfaces for the annulus and classify the phase diagrams by varying the inner and outer radii.
In both cases, we find the mutual information satisfies the monotonicity dictated by the unitarity and decays exponentially fast as the gap scale is increased.
We speculate this is a universal behavior in any gapped system.
}

\maketitle

\section{Introduction}
Entanglement entropy is an invaluable tool to explore various aspects of quantum field theories (QFTs) in diverse dimensions, such as 
critical phenomena \cite{Holzhey:1994we,Vidal:2002rm,Kitaev:2005dm,Levin:2006zz,Calabrese:2004eu,Ryu:2006bv,Ryu:2006ef,Grover:2011fa}, confinement/deconfinement phase transition \cite{Nishioka:2006gr,Klebanov:2007ws,Pakman:2008ui,Buividovich:2008gq,Nakagawa:2009jk} and renormalization group flow \cite{Casini:2004bw,Myers:2010xs,Myers:2010tj,Liu:2012eea,Casini:2012ei}. 
It depends on a state of interest and the shape of an entangling surface $\Sigma$ that divides a space into a region $A$ and its complement $\bar A$.
In the simplest case, $\Sigma$ is chosen to be a round sphere (or two endpoints of an interval in two dimensions) which allows us to use a conformal transformation \cite{Casini:2011kv,Calabrese:2004eu} and obtain analytic results for conformal field theories (CFTs).
Multiple disjoint intervals for CFT$_2$ are examined by \cite{Calabrese:2009ez,Calabrese:2010he,Headrick:2010zt,Hartman:2013mia,Faulkner:2013yia}, and small deformations of an entangling surface in CFT$_{d\ge 3}$ are perturbatively studied in \cite{Rosenhaus:2014woa,Rosenhaus:2014ula,Allais:2014ata,Lewkowycz:2014jia} recently.
More general shapes, however, have not been fully understood so far because of the computational complexity, especially in non-conformal field theories.\footnote{See e.g. \cite{Ben-Ami:2014gsa,Fischler:2012uv} for studies on non-conformal theories including finite temperature cases. 
}

A special case is a system with a large mass gap where entanglement entropy can be expanded in powers of the inverse of the gap.
The coefficients appearing in the expansion are unknown in general, but are assumed to be integrals of the functions of the extrinsic curvature and its derivatives of $\Sigma$ \cite{grover2011entanglement}.
The entanglement entropy for $\Sigma$ diffeomorphic to a circle is examined by \cite{Klebanov:2012yf} for free massive fields in three dimensions, showing that all the coefficients can be systematically determined by the logarithmic divergences of higher-dimensional theories that are the consequence of the conformal anomalies (see also \cite{Safdi:2012sn}).
Similar argument holds for $\Sigma$ diffeomorphic to multiple disjoint circles.

In this paper, we consider the entanglement entropy of an annulus in QFT$_3$ as a guide to investigate the phase structure of the ground state.\footnote{Refer to \cite{Sabella-Garnier:2014fda} as a related work on the fuzzy sphere.}
For CFT, the finite part of the entropy is a function of the ratio of the inner and outer radii $R_1 < R_2$.
Unfortunately, the conformal transformation used for a spherical entangling surface \cite{Casini:2011kv} does not help us identify the function for the annulus.
It, however, was shown in \cite{Hirata:2006jx} that the strong subadditivity \cite{lieb1973proof} requires the function is concave with respect to $\log (R_2/R_1)$.
In a gapped system, the finite part of the entropy depends not only on the ratio, but also on the gap scale, and thus there are no known constraints for the entropy from the strong subadditivity.

To check if the constraints from the strong subadditivity holds for CFT$_3$, or more generally to fix the dependence on the ratio and the gapped scale, 
we perform numerical calculations of the entanglement entropy for a free massive scalar field theory.
We put the scalar field on radial lattice following \cite{Srednicki:1993im} 
and compute the mutual information across the annulus (see Fig.\,\ref{fig:MI}).
The mutual information is better than entanglement entropy itself in a sense that it is free from UV divergences and independent of the regularization scheme.
For a massless scalar field, the strong subadditivity constrains the mutual information to be a convex function with respect to the ratio of the annulus. 
We confirm that the convexity holds in our results and inspect the limits of $R_2/R_1 \to 1$ and $R_2/R_1 \to \infty$.
In the former limit, we approximate the thin annulus by a thin strip (see Fig.\,\ref{fig:SmallWidth}) and evaluate the mutual information by dimensional reduction to an interval in $(1+1)$ dimensions.
In the latter case, we can conformally map the annulus to two disjoint circles whose entanglement entropy is studied both numerically and analytically in the large separation limit \cite{Shiba:2010dy,Shiba:2012np,Cardy:2013nua}.
We find that our fittings are consistent with the analytic results within our numerical precision. 
The implementation of the mass is straightforward numerically, and we observe that the mutual information exponentially decays as the mass increases while fixing the ratio $R_2/R_1$.

Another model we are able to tackle is a strongly coupled QFT holographically dual to the gravity on the AdS space.
The holographic calculation of entanglement entropy, known as the Ryu-Takayanagi formula \cite{Ryu:2006bv,Ryu:2006ef,Lewkowycz:2013nqa}, 
associates the given region $A$ in a QFT$_d$ to a codimension-two minimal surface $\gamma_A$ satisfying $\partial \gamma_A = \Sigma$ in the AdS$_{d+1}$ space, and gives the entropy $S_A$ by the area of the surface, $S_A = \text{Area}(\gamma_A)/(4G_N)$.
It can be applied to disjoint regions and exhibits interesting transitions between different minimal surfaces with the same boundary condition \cite{Headrick:2010zt}, each one of them corresponding to a specific phase in the dual QFT.

We study the holographic entanglement entropy of the annulus  by extending the work \cite{Klebanov:2012yf} for a disk in a confining gauge theory with a gap described by the CGLP background \cite{Cvetic:2000db}.
The entropy given by the area of a minimal surface \cite{Ryu:2006bv,Ryu:2006ef} shows a phase transition due to the change of the topology \cite{Liu:2012eea}.
There are four types of minimal surfaces anchored on the annulus, namely, (1) hemi-torus, (2) two disk, (3) two cylinder, and (4) one disk and one cylinder (disk-cylinder) types.
The last three solutions are superpositions of the disk- and cylinder-type solutions found in \cite{Klebanov:2012yf}.
The first one was also constructed by \cite{Hirata:2006jx} in the AdS space.
Since the mutual information vanishes for disconnected surfaces only the hemi-torus solution has a non-zero value.
Comparing their areas we classify the four phases in the $(R_1, R_2)$-plane as shown in Fig.\,\ref{fig:PhaseDiagram}.
The holographic model also exhibits the exponential decay of the mutual information with respect to the gap with the ratio $R_2/R_1$ fixed.

Based on the observations in the free massive scalar and the holographic models,  we speculate that the mutual information through an annulus decays exponentially as
\begin{align}\label{MI_Speculation}
        I_\text{annulus} \sim \exp[- \# m(R_2 - R_1)] \ ,
\end{align}
in any gapped system with a gap scale $m$.

\section{Entanglement entropy of annulus}\label{sec:EEofAnnulus}
In this section, we review general properties of the entanglement entropy for an annulus $A$ in CFT and a gapped system.
We discuss its relation to the mutual information between the inner disk and the compliment of the outer disk.

\subsection{Conformal field theory}
In a three-dimensional CFT,
the form of the entropy for an annulus $A$ is fixed by the conformal symmetry,
\begin{align}\label{S_A_CFT}
S_A(R_1, R_2) = \alpha \frac{2\pi (R_1 + R_2)}{\epsilon}-f(R_2/R_1) \ ,
\end{align}
where the first term obeys the area law with the UV cutoff length $\epsilon$ and the second term $f$ is a function of the ratio $R_2/R_1$ of the radii. 
This function $f$ should be monotonically decreasing and convex
\begin{align}\label{SSA}
f'(\rho) \le 0 \ , \qquad f''(\rho) \ge 0 \ ,
\end{align}
with respect to the new variable $\rho = \log (R_2/R_1)$, due to the strong subadditivity
\begin{align}\label{SSA_def}
S_B+S_C\ge S_{B\cup C}+S_{B\cap C} \ .
\end{align}
In what follows, we review the derivation of \eqref{SSA} given by \cite{Hirata:2006jx}.

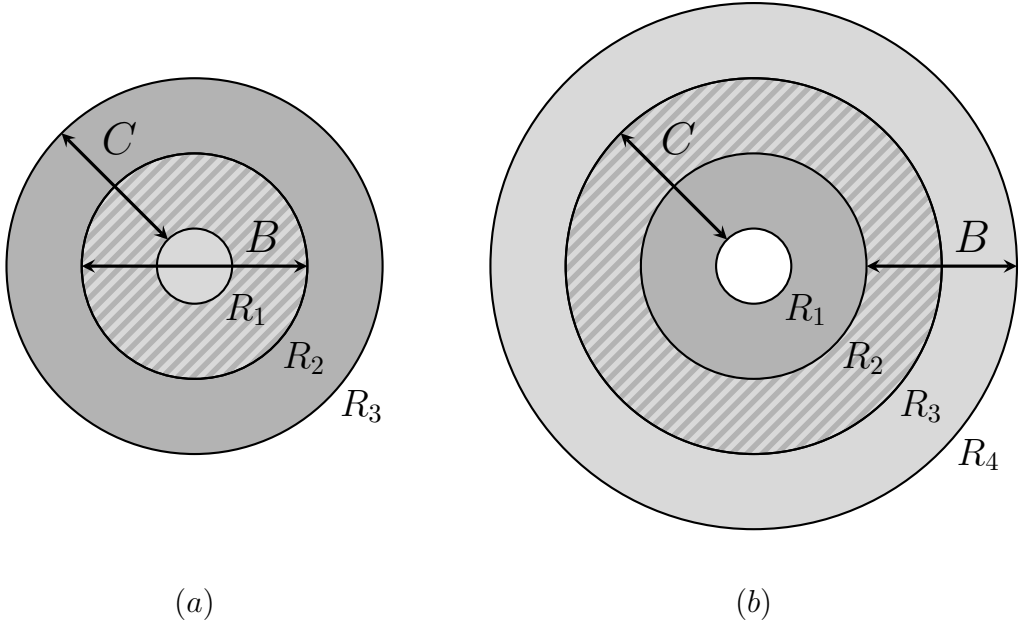
\begin{figure}[htbp]
\centering
\begin{tikzpicture}[thick, >=stealth]

    \pgfdeclarepatternformonly{swnestripes}{\pgfpoint{0cm}{0cm}}{\pgfpoint{1cm}{1cm}}{\pgfpoint{1cm}{1cm}}
    {
        \foreach \i in {0.1cm, 0.3cm,...,0.9cm}
        {
         \pgfpathmoveto{\pgfpoint{\i}{0cm}}
         \pgfpathlineto{\pgfpoint{1cm}{1cm - \i}}
         \pgfpathlineto{\pgfpoint{1cm}{1cm - \i + 0.1cm}}
         \pgfpathlineto{\pgfpoint{\i - 0.1cm}{0cm}}
         \pgfpathclose%
         \pgfusepath{fill}
         \pgfpathmoveto{\pgfpoint{0cm}{\i}}
         \pgfpathlineto{\pgfpoint{1cm - \i}{1cm}}
         \pgfpathlineto{\pgfpoint{1cm - \i - 0.1cm}{1cm}}
         \pgfpathlineto{\pgfpoint{0cm}{\i + 0.1cm}}
         \pgfpathclose%
         \pgfusepath{fill}
        }
    }
    \pgfdeclarepatternformonly{swneStripes}{\pgfpoint{0cm}{0cm}}{\pgfpoint{1cm}{1cm}}{\pgfpoint{1cm}{1cm}}
    {
        \foreach \i in {0.1cm, 0.3cm,...,0.9cm}
        {
         \pgfpathmoveto{\pgfpoint{\i}{0cm}}
         \pgfpathlineto{\pgfpoint{1cm}{1cm - \i}}
         \pgfpathlineto{\pgfpoint{1cm}{1cm - \i - 0.1cm}}
         \pgfpathlineto{\pgfpoint{\i + 0.1cm}{0cm}}
         \pgfpathclose%
         \pgfusepath{fill}
         \pgfpathmoveto{\pgfpoint{0cm}{\i}}
         \pgfpathlineto{\pgfpoint{1cm - \i}{1cm}}
         \pgfpathlineto{\pgfpoint{1cm - \i + 0.1cm}{1cm}}
         \pgfpathlineto{\pgfpoint{0cm}{\i - 0.1cm}}
         \pgfpathclose%
         \pgfusepath{fill}
        }
    }
    \pgfdeclarepatternformonly{senwstripes}{\pgfpoint{0cm}{0cm}}{\pgfpoint{1cm}{1cm}}{\pgfpoint{1cm}{1cm}}
    {
        \foreach \i in {0.1cm, 0.3cm,...,0.9cm}
        {
         \pgfpathmoveto{\pgfpoint{0cm}{\i}}
         \pgfpathlineto{\pgfpoint{0cm}{\i + 0.1cm}}
         \pgfpathlineto{\pgfpoint{\i + 0.1cm}{0cm}}
         \pgfpathlineto{\pgfpoint{\i}{0cm}}
         \pgfpathclose%
         \pgfusepath{fill}
         \pgfpathmoveto{\pgfpoint{1cm}{\i}}
         \pgfpathlineto{\pgfpoint{\i}{1cm}}
         \pgfpathlineto{\pgfpoint{\i + 0.1cm}{1cm}}
         \pgfpathlineto{\pgfpoint{1cm}{\i + 0.1cm}}
         \pgfpathclose%
         \pgfusepath{fill}
        }
    }
    \pgfdeclarepatternformonly{senwStripes}{\pgfpoint{0cm}{0cm}}{\pgfpoint{1cm}{1cm}}{\pgfpoint{1cm}{1cm}}
    {
        \foreach \i in {0.0cm, 0.2cm,...,0.8cm}
        {
         \pgfpathmoveto{\pgfpoint{\i}{0cm}}
         \pgfpathlineto{\pgfpoint{0cm}{\i}}
         \pgfpathlineto{\pgfpoint{0cm}{\i + 0.1cm}}
         \pgfpathlineto{\pgfpoint{\i + 0.1cm}{0cm}}
         \pgfpathclose%
         \pgfusepath{fill}
         \pgfpathmoveto{\pgfpoint{1cm}{\i}}
         \pgfpathlineto{\pgfpoint{\i}{1cm}}
         \pgfpathlineto{\pgfpoint{\i + 0.1cm}{1cm}}
         \pgfpathlineto{\pgfpoint{1cm}{\i + 0.1cm}}
         \pgfpathclose%
         \pgfusepath{fill}
        }
    }
        \draw[fill=black!30] (0,0) circle [radius=2.5];
        \filldraw[pattern=swnestripes,pattern color=black!15] (0,0) circle [radius=1.5];
        \draw[] (0,0) circle [radius=1.5];
        \draw[fill=black!15] (0,0) circle [radius=0.5];
        \draw [very thick, <->] (180:3/2) -- (0:3/2);
        \draw (0.9,0.4) node[] {\Large $B$};
        \draw [very thick, <->] (135:1/2) --++ (135:2);
        \draw (-1,1.7) node[] {\Large $C$};
        \draw (-40:0.9) node {\large $R_1$};
        \draw (-40:1.9) node {\large $R_2$};
        \draw (-40:2.9) node {\large $R_3$};
        \draw (0,-9/2) node {($a$)};
\end{tikzpicture}
\hspace{1 cm}
\begin{tikzpicture}[thick, >=stealth]
        \draw[fill=black!15] (0,0) circle [radius=3.5];
        \filldraw[pattern=swnestripes,pattern color=black!30] (0,0) circle [radius=2.5];
        \draw[] (0,0) circle [radius=2.5];
        \draw[fill=black!30] (0,0) circle [radius=1.5];
        \draw[fill=white] (0,0) circle [radius=0.5];
        \draw [very thick, <->] (0:3/2) -- (0:7/2);
        \draw (2.9,0.4) node[] {\Large $B$};
        \draw [very thick, <->] (135:1/2) --++ (135:2);
        \draw (-1,1.7) node[] {\Large $C$};
        \draw (-40:0.9) node {\large $R_1$};
        \draw (-40:1.9) node {\large $R_2$};
        \draw (-40:2.9) node {\large $R_3$};
        \draw (-40:3.9) node {\large $R_4$};
        \draw (0,-9/2) node {($b$)};
\end{tikzpicture}

\caption{The subsystems $B$ (in light gray) and $C$ (in dark gray) to prove the monotonicity ($a$) and the convexity ($b$) of the function $f$ in \eqref{S_A_CFT}. 
The striped regions are the intersections $B\cap C$.}
\label{fig:SSA}
\end{figure}

Let the regions $B$ and $C$ be a disk of radius $R_2$ and an annulus of radii $R_1$ and $R_3$ with $R_1<R_2<R_3$ as in Fig.\,\ref{fig:SSA} $(a)$. 
The entanglement entropy for a disk of radius $R$ takes a form of
\begin{align}\label{S_disk}
  S_\text{disk}(R)=\alpha\frac{2\pi R}{\epsilon} - F \ ,
\end{align}
with a constant $F$.\footnote{The constant equals to the free energy on $S^3$, $F = -\log Z(S^3)$ \cite{Casini:2011kv}.} 
The strong subadditivity \eqref{SSA_def} together with \eqref{S_A_CFT} and \eqref{S_disk} yields the monotonicity in \eqref{SSA}:
\begin{align}
  f(R_3/R_1)\le f(R_2/R_1)\ .
\end{align}
The convexity in \eqref{SSA} can be derived similarly by taking both $B$ and $C$ as an annulus of radii $R_2$ and $R_4$, and an annulus of radii $R_1$ and $R_3$ satisfying $R_1<R_2<R_3<R_4$ as in Fig.\,\ref{fig:SSA} $(b)$.
In the $R_4\to R_3$ limit, the strong subadditivity
\begin{align}
  f(R_4/R_2)+f(R_3/R_1)\le f(R_4/R_1)+f(R_3/R_2) \ ,
\end{align}
reduces to the monotonicity of $f'(\rho)$.

\subsection{A gapped system}
In theories with a mass gap of order $m$, the entanglement entropy of a region $A$ has an expansion in powers of $1/m$:
\begin{align}\label{S_Gapped}
S_A = \alpha \frac{\ell_\Sigma}{\epsilon} + \beta\, m\,\ell_\Sigma - \gamma_\Sigma + \sum_{n=0}^\infty\frac{c_{2n+1}^\Sigma}{m^{2n+1}} \ ,
\end{align}
with numerical constants $\alpha,\beta$ and the topological entanglement entropy  $\gamma_\Sigma$ \cite{Kitaev:2005dm,Levin:2006zz}. 
Here the $\gamma_\Sigma$ depends on only the topology of the entangling surface $\Sigma = \partial A$ and detects long-range order. 
The dimensionful coefficients $c_{2n+1}^\Sigma$ are postulated \cite{grover2011entanglement} as local integrals 
of functions of the extrinsic curvature and its derivatives on $\Sigma$. 
In other words, entanglement contributing to $c_{2n+1}^\Sigma$ localizes on the entangling surface in the large-$m$ limit due to the short correlation length of order $1/m$.

Applying \eqref{S_Gapped} to the annulus $A$ of our interest, the entanglement entropy should take the form of
\begin{align}\label{S_A_Gapped}
\begin{aligned}
S_A (R_1, R_2,m) &= \alpha \frac{2\pi (R_1 + R_2)}{\epsilon} + 2\pi \beta\, m\, (R_1 + R_2)- \gamma_{\Sigma} + \sum_{n=0}^\infty\frac{c_{2n+1}^\Sigma}{m^{2n+1}}\ ,
\end{aligned}
\end{align}
where $\Sigma$ is two concentric circles of radii $R_1$ and $R_2$.
The coefficients $c_{2n+1}^\Sigma$ are polynomials of the radii of order $-(2n+1)$.

\subsection{Mutual information}
The mutual information between two disjoint regions $B$ and $C$ is defined out of the entanglement entropies as
\begin{align}
I(B,C) \equiv S_B + S_C - S_{B\cup C} \ .
\end{align}
It is always finite because the area law divergences cancel by definition, and non-negative because of the subadditivity $S_{B\cup C}\le S_B+S_C$.
In addition, the strong subadditivity yields the monotonicity of the mutual information
\begin{align}\label{Monotonicity_MI}
        I(B,C) \le I (B,C \cup D) \ , 
\end{align}
for any region $D$.

\begin{figure}[htbp]
\centering
\begin{tikzpicture}[thick, >=stealth]
        \draw [fill=red!20] (0,0) circle (3cm);
        \draw [fill=white] (0,0) circle (1.5cm);
        \draw [very thick, ->] (0,0) --++ (0:1.5cm);
        \draw [very thick, ->] (0,0) -- ++(30:3cm);
        \node at (-0.5,0) {\Large $B$};
        \node at (-4,0) {\Large $C$};
        \node at (0,-2) {\Large $A=\overline{B\cup C}$};
        \node at (0.75,-0.5) {\Large$R_1$};
        \node at (1.5,1.5) {\Large$R_2$};
\end{tikzpicture}
\caption{The entangling regions for the mutual information. The region $B$ is a disk of radius $R_1$. The region $C$ is the complement of a disk of radius $R_2$. The complement of the union of the two regions $\overline{B\cup C}$ is the annulus $A$ in the red colored region.}
\label{fig:MI}
\end{figure}
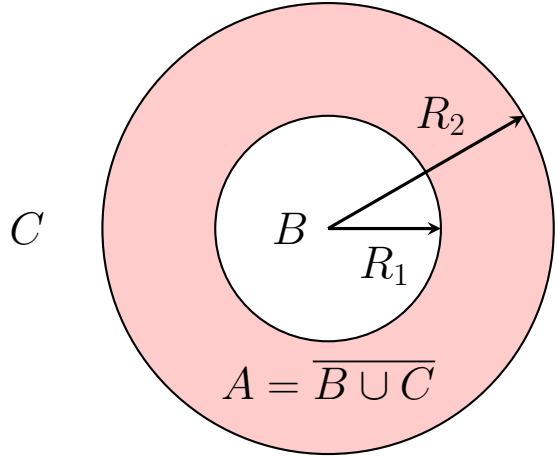

To extract the finite parts of the entanglement entropies \eqref{S_A_CFT} and \eqref{S_A_Gapped} of the annulus $A$, we take $B$ and $C$ to be two regions outside $A$, namely, a disk of radius $R_1$ and the complement of a disk of radius $R_2$, respectively (see Fig.\,\ref{fig:MI}).
We can interpret this mutual information as how much quantum information is shared by $B$ and $C$ across the annulus $A$. 
Since the entanglement entropy of a given region is equal to that of the complement, $S_C$ and $S_{B\cup C}$ equal the entropies of a disk of radius $R_2$ and an annulus of inner and outer radii $R_1$ and $R_2$, respectively.
The mutual information $I$ across the annulus $A$ then reduces to 
\begin{align}\label{MI}
  I(R_1,R_2)\equiv I(B,C)=S_\text{disk}(R_1)+S_\text{disk}(R_2)-S_A(R_1,R_2) \ .
\end{align}

In this setup, the inequality \eqref{Monotonicity_MI} translates into the monotonicity of $I$ with respect to the radii $R_1, R_2$:
\begin{align}
\begin{aligned}\label{MI_Monotonicity}
  \frac{\partial}{\partial R_1}I(R_1,R_2)\ge 0\ , \qquad
  \frac{\partial}{\partial R_2}I(R_1,R_2)\le 0 \ ,
\end{aligned}
\end{align}
which holds for any unitary QFT.
The proof proceeds as follows: Let $B,C$ be the regions in Fig.\,\ref{fig:MI} and $D$ be an annulus of radii $R_1+\Delta R_1$ and $R_2$, then the monotonicity $I(B,C)\le I(B\cup D,C)$ yields $I(R_1,R_2)\le I(R_1+\Delta R_1,R_2)$.
Similarly let $D$ be an annulus of radii $R_1$ and $R_2-\Delta R$, then the monotonicity $I(B,C)\le I(B,C\cup D)$ yields $I(R_1,R_2)\le I(R_1,R_2-\Delta R_2)$.

For CFT, the mutual information becomes
\begin{align}\label{I_CFT}
I_\text{CFT} = f(R_2/R_1)-2F \ ,
\end{align}
with the constant $F$ in the disk entropy \eqref{S_disk}, and the inequalities \eqref{MI_Monotonicity} are equivalent to the monotonicity of $f$ that was already derived in \eqref{SSA}.

On the other hand, applying \eqref{S_A_Gapped} for a gapped system to \eqref{MI} leads to
\begin{align}\label{I_Gapped}
I_\text{gapped} = \gamma_\Sigma-2\gamma_\text{disk} \ .
\end{align}
Here the $m$-dependent terms cancel out due to the assumption that the dimensionful coefficients $c_{2n+1}^\Sigma$ in \eqref{S_A_Gapped} are  integrals on the entangling region $\Sigma$. 
Note that the expression \eqref{I_Gapped} would fail for small masses such as $mR_1\lesssim 1$ or $m(R_2-R_1)\lesssim 1$ if there could exist an exponential term like $O\left( \exp [ - \# m ] \right)$ to \eqref{S_Gapped} which can not be seen in the large mass expansion.
We will discuss such a correction in Section \ref{ss:Universal}.

In the following sections,
we will use these mutual informations \eqref{I_CFT} and \eqref{I_Gapped} to determine the function $f$ in CFT and to check whether the large mass expansion formula \eqref{S_Gapped} holds for the annulus.

\section{Free massive scalar field}\label{ss:FreeMassiveScalar}

Let us apply the general discussion on the annulus entropy in Section 2 to a free massive scalar field whose action is defined by
\begin{align}\label{ScalarAction}
I = \frac{1}{2}\int d^3 x \left[ (\partial_\mu \phi)^2 + m^2 \phi^2 \right] \ .
\end{align}
In this case, the coefficients $\beta$ and $\gamma$ in the entropy \eqref{S_A_Gapped} are known to be $\beta=-1/12$ and $\gamma=0$.\footnote{The topological entanglement entropy vanishes because there is an empty theory in the IR of the massive scalar theory.}
The coefficients $c^\Sigma_{2n+1}$ are  calculated \cite{Klebanov:2012yf,Safdi:2012sn} up to $n=1$, being local integrals 
of functions of the extrinsic curvature $\kappa$ and $\kappa$'s derivatives on the $\Sigma$.
For example,
\begin{align}
  c_1=-\frac{n_0+3n_{1/2}}{480}\int_\Sigma ds\,\kappa^2 \ ,
\end{align}
for $n_0$ free scalar fields and $n_{1/2}$ free Dirac fermions.
In the present case, the entangling surface is two disjoint disks of radii $R_1,R_2$ whose extrinsic curvatures are $\kappa = 1/R_1, 1/R_2$.
Thus $c_1^\Sigma=-\frac{\pi}{240}(1/R_1+1/R_2)$ for a single free scalar field.
The constant term $F$ of the disk entropy \eqref{S_disk} is analytically calculated  as the free energy on a three-sphere, $F_\text{scalar}=(\ln 2)/8-3\zeta(3)/16\pi^2\simeq0.0638$ \cite{Klebanov:2011gs}.
The mutual informations \eqref{I_CFT} and \eqref{I_Gapped} are 
\begin{align}\label{I_massless}
  I_\text{massless}&=f(R_2/R_1)-2F_\text{scalar}\ ,\\
  I_\text{massive}&=0 \ . \label{I_massive}
\end{align}

\subsection{Numerical results}

\begin{figure}[htbp]
  \includegraphics[width=8cm]{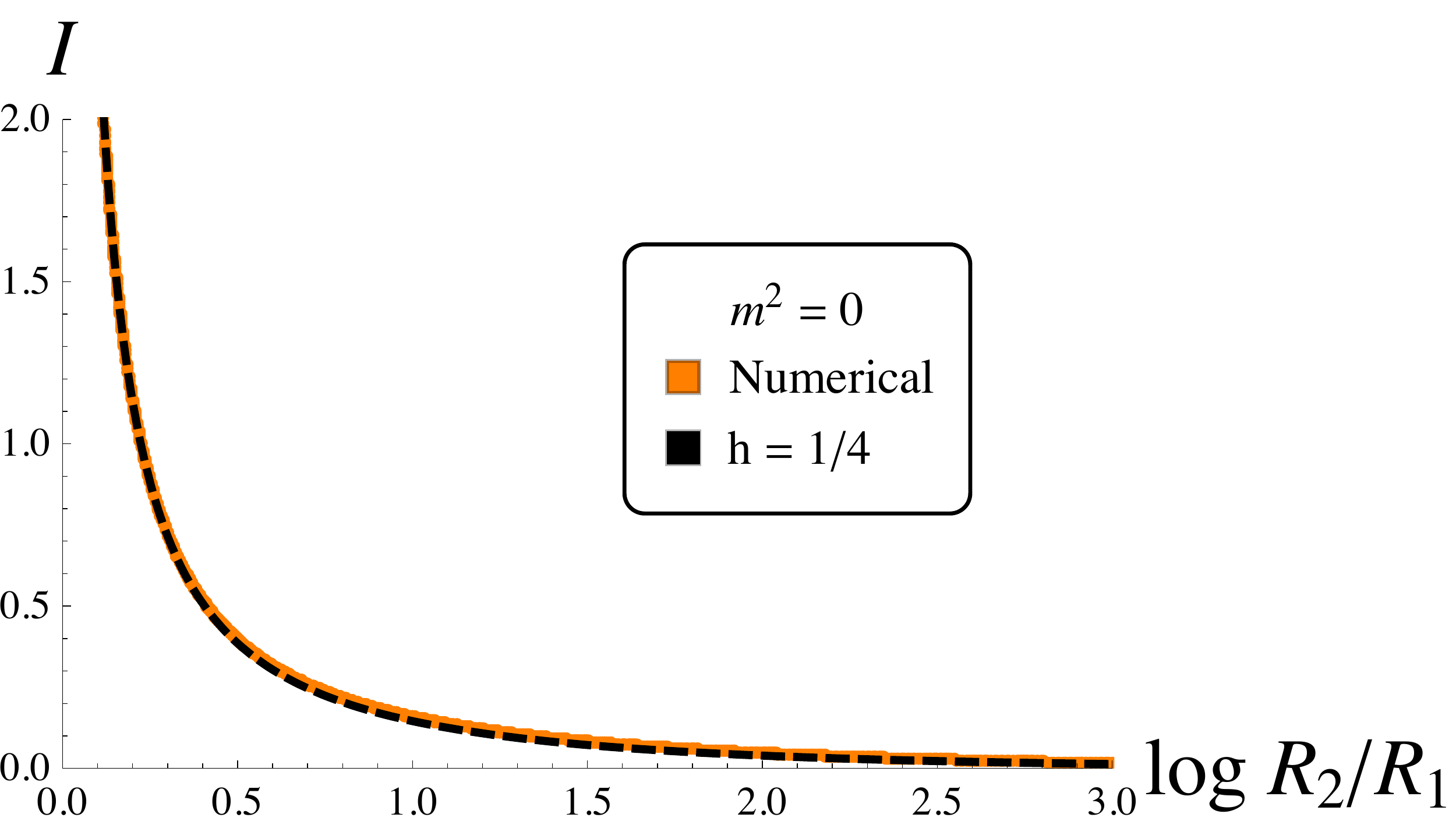}
  \includegraphics[width=9cm]{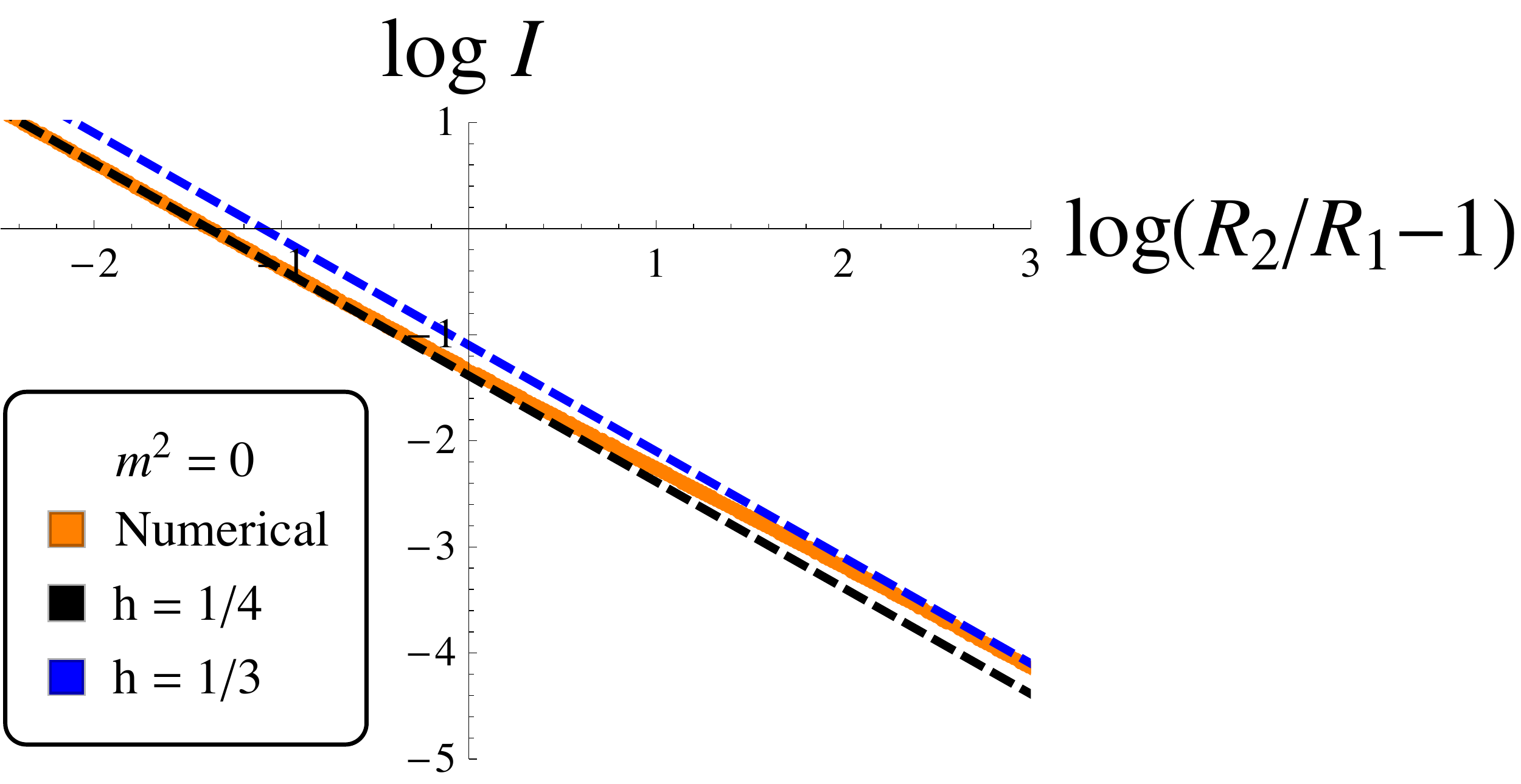}
\begin{center}
   ($a$)\hspace{7cm} ($b$)
\end{center}
\caption{The mutual informations $I$ across the annulus of radii $R_1$ and $R_2$ for the free massless scalar field. ($a$) The mutual information $I$ (the orange line) has the desired monotonicity and convexity, and is well fitted by $h/(R_2/R_1-1)$ (the black dotted line). ($b$) However, this coefficient $h$ is not a constant and increases with $R_2/R_1$ from $h\simeq 1/4$ (the black dotted line) to $h\simeq 1/3$ (the blue dotted line).}
\label{fig:masslessI}
\end{figure}

\begin{figure}[htbp]
  \includegraphics[width=8cm]{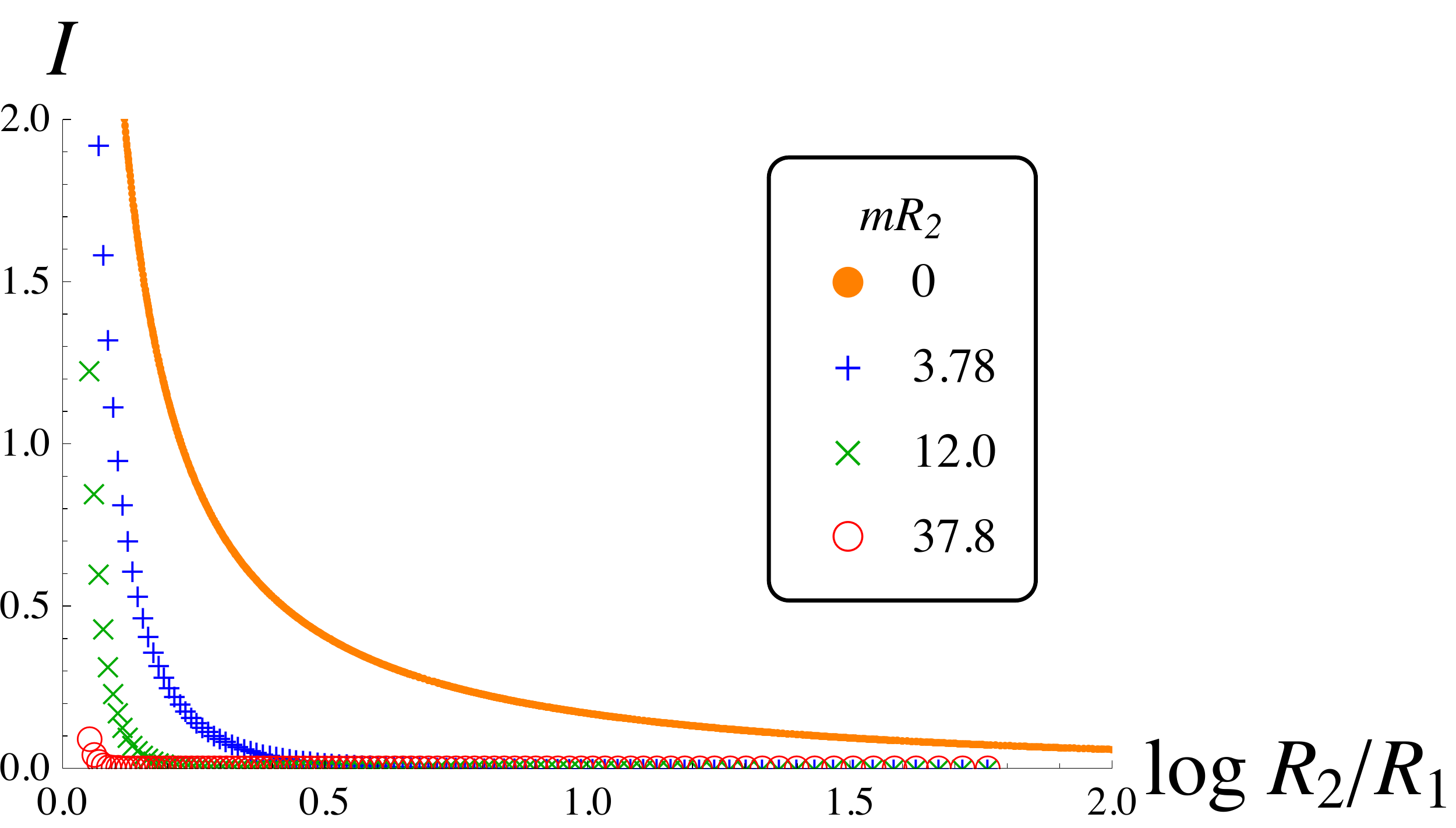}
  \includegraphics[width=8cm]{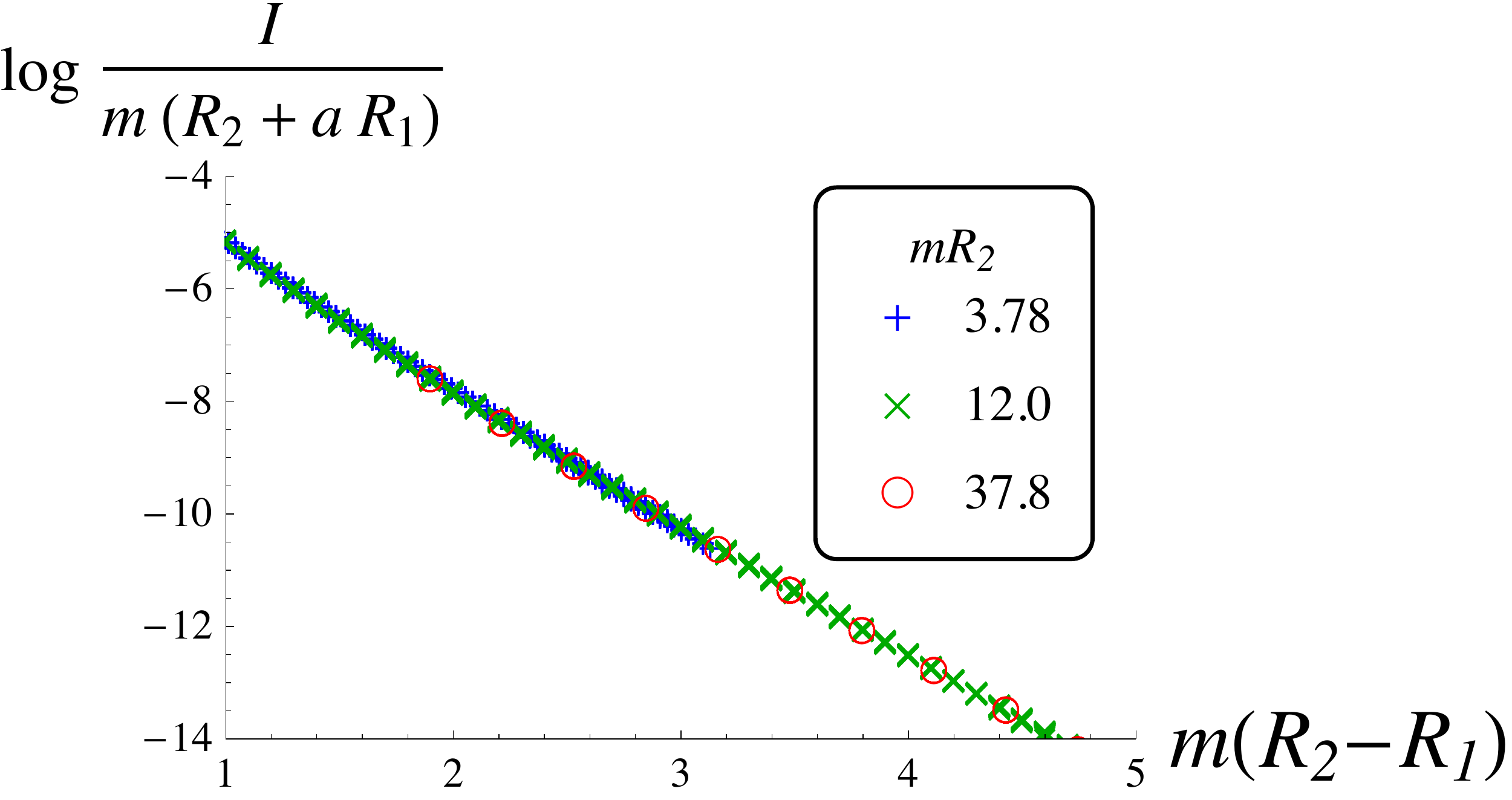}
\begin{center}
   ($a$)\hspace{7cm} ($b$)
\end{center}
\caption{The mutual informations $I$ across the annulus of radii $R_1$ and $R_2$ for scalar fields with different masses $m$. 
($a$) $I$ monotonically decreases with the mass $m$ (orange$\to$blue$\to$green$\to$red). 
($b$) In fact, it exponentially decreases with the dimensionless width $m(R_2-R_1)$. 
For $m(R_2-R_1)\gtrsim1$, it shows $I\propto m(R_2+a R_1)\exp[-b\,m(R_2-R_1)]$ with $a\simeq 2\sim5$ ($a=3$ in the figure) and $b\simeq 2.5$. }
\label{fig:massiveI}
\end{figure}

We perform the numerical calculation by putting a free scalar field on the radial lattice following \cite{Srednicki:1993im,Huerta:2011qi}, whose
 details can be found in Appendix \ref{ss:Numerics}.
The main results are presented in Fig.\,\ref{fig:masslessI} and \ref{fig:massiveI}. 

Fig.\,\ref{fig:masslessI} shows
the mutual information $I$ \eqref{I_massless} for the free massless scalar field. 
The function $f$ satisfies the desired monotonicity and convexity \eqref{SSA} with respect to $\rho=\log(R_2/R_1)$ as is clear in Fig.\,\ref{fig:masslessI} ($a$). 
The mutual information $I$ asymptotically vanishes for large $R_2/R_1$, which means that the function $f$ has a finite constant term  $2F_\text{scalar}$. 
This suggests that the finite constant term is topological and additive for each connected component of the entangling surfaces, namely, proportional to the 0-th Betti number $b_0[\Sigma]$ of $\Sigma$. 
The numerical function $I=I(R_2/R_1)$ is well approximated by $h/(R_2/R_1-1)$ with $h\simeq1/4$ for small width, but $h$ monotonically increases to $h\simeq1/3$ for large $R_2/R_1$ (see Fig.\,\ref{fig:masslessI} ($b$)).  
We therefore propose that $f$ is given by 
\begin{align}\label{fCFT}
  f(R_2/R_1)=\frac{h(R_2/R_1)}{R_2/R_1-1}+2F_\text{scalar} \ ,
\end{align}
where $h(R_2/R_1)$ is a mild monotonically increasing function of $R_2/R_1$ such that $h\simeq1/4$ for $R_2/R_1\sim1$ and $h\simeq1/3$ for $R_2/R_1\gg1$. 
These asymptotic values are consistent with previous works \cite{Casini:2005zv,Cardy:2013nua} as will be explained in the next subsections.

The mutual information \eqref{I_massless} for the free massive scalar field is displayed in Fig.\,\ref{fig:massiveI}.
It is monotonically decreasing with the mass (i.e., decreasing with $mR_2$ or $mR_1$ while $R_2/R_1$ being fixed), and almost vanishes for large mass (Fig.\,\ref{fig:massiveI} ($a$)) as is consistent with \eqref{I_massive}.
In fact, Fig.\,\ref{fig:massiveI} ($b$) demonstrates that the mutual information decays exponentially with a ``dimensionless width'' $m(R_2-R_1)$,
\begin{align}\label{I_ExpDecay}
  I_\text{massive}\propto m(R_2+aR_1) \exp[-b\, m(R_2-R_1)] \ ,
\end{align}
with constants $a$ and $b$.
This exponential behaviour satisfies the expected monotonicity \eqref{MI_Monotonicity}.
We will find similar decay even in the holographic model in Section \ref{ss:HEE} and discuss their possible universality in a gapped phase in Section \ref{ss:Universal}.

\subsection{Small and large width limits in CFT}
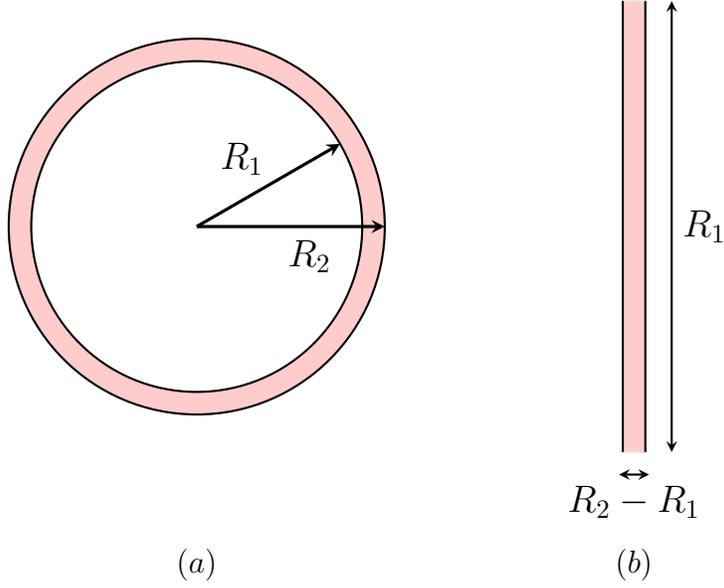
\begin{figure}[htbp]
\centering
\begin{tikzpicture}[thick, >=stealth]
        \draw[fill=red!20] (0,0) circle [radius=2.5];
        \draw[fill=white] (0,0) circle [radius=2.2];
        \draw [very thick, ->] (0,0) -- (30:2.2);
        \draw [very thick, ->] (0,0) -- (5/2,0);
        \draw (0.6,0.9) node {\large $R_1$};
        \draw (1.5,-0.4) node {\large $R_2$};
        \draw (0,-9/2) node {($a$)};
\end{tikzpicture}
\hspace{2 cm}
\begin{tikzpicture}[thick, >=stealth]
        \draw[gray!20,fill=red!20] (-0.15,-3) --++ (0.3, 0) --++ (0,6) --++ (-0.3,0) --++ (0,-6);
        \draw (-0.15,-3) --++ (0,6);
        \draw (0.15,-3) --++ (0,6);
        \draw[<->] (0.5,-3) --++ (0,6) node[midway, right] {\large $R_1$};
        \draw[<->] (-0.2,-3.3) --++  (0.4,0) node[midway, below] {\large $R_2 - R_1$};
        \draw (0,-9/2) node {($b$)};
\end{tikzpicture}

\caption{A thin annulus ($a$) can be approximated by a thin strip ($b$) with compactified direction.}
\label{fig:SmallWidth}
\end{figure}

The annulus with small width ($R_2/R_1 \approx 1$) can be approximated by a thin strip of width $R_2 - R_1$ extending along a circle of radius $2\pi R_1$ as in Fig.\,\ref{fig:SmallWidth}.\footnote{We thank T.\,Takayanagi for drawing our attention to this point.}
The mutual information for the thin strip of width $\delta$ is shown to obey \cite{Casini:2005zv,Casini:2009sr}
\begin{align}\label{ThinMI}
        I \simeq \kappa\, \frac{\CA}{\delta} \ ,
\end{align}
where $\CA$ is the area of the plane bounding the strip.
This behavior was derived by dimensionally reducing the thin strip to an interval in $(1+1)$ dimensions for free fields and summing the mutual informations over the Kaluza-Klein modes.

The coefficient $\kappa$ is calculated for a free massless scalar field \cite{Casini:2005zv} to be $\kappa = 0.0397$. 
Applying \eqref{ThinMI} to our case, we find
\begin{align}\label{SmallWidthLImit}
        I \simeq 0.0397 \, \frac{2\pi R_1}{R_2 - R_1} = \frac{0.249}{R_2/R_1 - 1} \ ,
\end{align}
which fits our numerical result in the small width limit ($h\simeq 1/4$ in \eqref{fCFT}) very well.
One may wonder if the small width limit of the mutual information \eqref{ThinMI} is universal and the coefficient $\kappa$ counts the number of degrees of freedom in any QFT.
We will come back to this point in Section \ref{ss:Universal} where we calculate $\kappa$ in a holographic model.

Next, consider the opposite limit where the width is large.
Let $w_i, z_i~(i=1,2)$ be the two-dimensional Cartesian coordinates related by an inversion transformation 
\begin{align}\label{InversionMap}
(z - z_0)_i = R_T^2 \frac{(w - w_0)_i}{|w - w_0|^2} \ ,
\end{align}
where $w_0$ is the inversion point. 
The inverse map is obtained by exchanging the role of $w$ and $z$ in the  transformation with the inversion point at $z=z_0$.
$R_T$ is a constant which we can tune arbitrarily. 

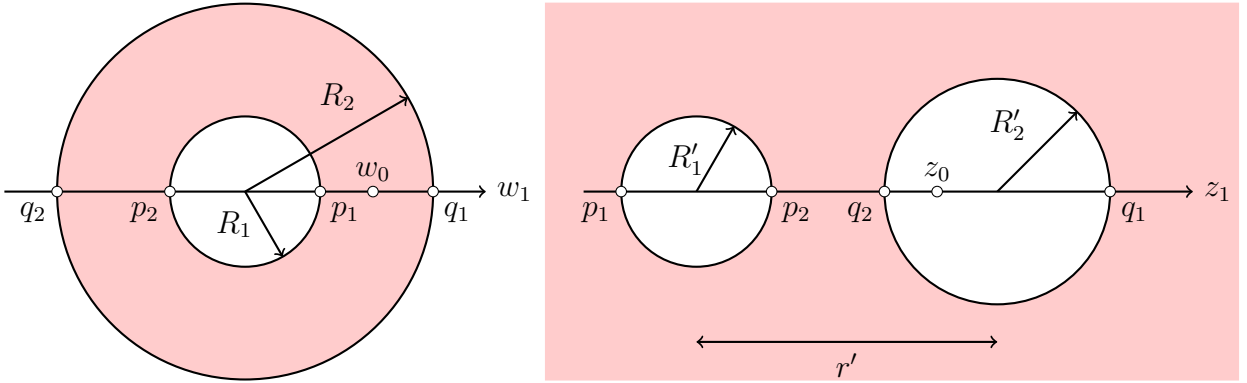
\begin{figure}[htbp]
\centering
\begin{tikzpicture}
        \begin{scope}[shift={(-10,0)}]
        \draw [name path=circleone, thick, fill=red!20] (0,0) circle (2.5cm);
        \draw [name path=circletwo, thick, fill=white] (0,0) circle (1cm);
        \draw [thick, ->] (0,0) -- ++(-60:1cm) node[midway, left] {$R_1$};
        \draw [thick, ->] (0,0) -- ++(30:2.5cm) node[near end, above left] {$R_2$};
        \draw [name path=line, thick, ->, fill=white] (-3.2,0) -- (3.2,0) node[right] {$w_1$};
        \draw [name intersections={of=circleone and line, by={q1, q2}}, fill=white] 
        (q1) circle (2pt) node [below right] at (q1) {$q_1$}
        (q2) circle (2pt) node [below left] at (q2) {$q_2$};
        \draw [name intersections={of=circletwo and line, by={p1, p2}}, fill=white] 
        (p1) circle (2pt) node [below right] at (p1) {$p_1$}
        (p2) circle (2pt) node [below left] at (p2) {$p_2$};
        \draw[fill=white] (1.7, 0) circle (2pt) node[above] {$w_0$};
        \end{scope}
        \begin{scope}[shift={(-3,0)}]
        \draw [thick, fill=red!20, red!20] (-3, -2.5) rectangle (6.2, 2.5);
        \draw [name path=circletwo, thick, fill=white] (-1,0) circle (1cm);
        \draw [name path=circleone, thick, fill=white] (3,0) circle (1.5cm);
        \draw [name path=line, thick, ->, fill=white] (-2.5,0) -- (5.6,0) node[right] {$z_1$};
        \draw [thick, ->] (-1,0) -- ++(60:1cm) node[midway, left] {$R_1'$};
        \draw [thick, ->] (3,0) -- ++(45:1.5cm) node[midway, above left] {$R_2'$};
        \draw [name intersections={of=circleone and line, by={q1, q2}}, fill=white] 
        (q1) circle (2pt) node [below right] at (q1) {$q_1$}
        (q2) circle (2pt) node [below left] at (q2) {$q_2$};
        \draw [name intersections={of=circletwo and line, by={p2, p1}}, fill=white] 
        (p1) circle (2pt) node [below left] at (p1) {$p_1$}
        (p2) circle (2pt) node [below right] at (p2) {$p_2$};
        \draw[fill=white] (2.2, 0) circle (2pt) node[above] {$z_0$};
        \draw[thick, <->] (-1, -2) -- (3, -2) node[midway, below] {$r'$};
        \end{scope}
\end{tikzpicture}
\caption{The inversion map of an annulus to two disjoint circles. The region inside the annulus in red color is mapped to the outside of the two circles in red color.}
\label{fig:MapToCircles}
\end{figure}

Consider an annulus in the $w$-coordinates whose center is at the origin with radii $R_1 < R_2$.
Let the points at $w_2=0$ on the outer circle be $p_1, p_2$ and on the inner circle be $q_1, q_2$.
We choose the inversion points $w_0$ and $z_0$ on the real axes at $(w_1, w_2)=(R_0, 0)$ and $(z_1,z_2) = (R_0', 0)$, respectively.
We assume $w_0$ is inside the annulus, $R_1 < R_0 < R_2$.
Under the transformation \eqref{InversionMap}, the annulus is mapped to two disjoint circles\footnote{We thank K.\,Ohmori and Y.\,Tachikawa for the discussions on this map.} (see Fig.\,\ref{fig:MapToCircles}) and the points $p_1, p_2$ and $q_1, q_2$ are at the intersections of the real axis and circles of radii $R_1'$ and $R_2'$ given by
\begin{align}\label{InversionMap1}
R_1' =R_T^2 \frac{R_1}{R_0^2 - R_1^2} \ , \qquad R_2' =R_T^2 \frac{R_2}{R_2^2 - R_0^2} \ .
\end{align}
The distance between the centers of the two circles is 
\begin{align}\label{InversionMap2}
r' = R_T^2 R_0 \frac{R_2^2 - R_1^2}{(R_2^2 - R_0^2) (R_0^2 - R_1^2)} \ .
\end{align}
The conformal symmetry implies that the cross ratio\footnote{
There are two cross ratios for four points.
The other one is
\begin{align*}
  y=\frac{|p_1 - p_2| |q_1 - q_2|}{|p_1 - q_1| |p_2 - q_2|}\ .
\end{align*}
} $x$ is invariant under the conformal transformation,
\begin{align}
x = \frac{|p_1 - p_2| |q_1 - q_2|}{|p_1 - q_2| |p_2 - q_1|} \ ,
\end{align}
which in our case is
\begin{align}
x = \frac{4R_1 R_2}{(R_1 + R_2)^2} = \frac{4R_1' R_2'}{r'^2 - (R_1' - R_2')^2} \ .
\end{align}

In this way, we can calculate the entanglement entropy of two disjoint circles from that of the corresponding annulus.
The former was studied in \cite{Shiba:2010dy,Shiba:2012np,Cardy:2013nua} in the widely separated limit for a free massless scalar field.
The mutual information between the two circles is \cite{Cardy:2013nua} 
\begin{align}
I = \frac{1}{3}\frac{R_1' R_2'}{r'^2} + O\left( (R_1' R_2'/r'^2)^2\right) \ .
\end{align}
The inversion maps \eqref{InversionMap1} and \eqref{InversionMap2} convert it to the mutual information of the annulus,
\begin{align}
I = \frac{1}{3} \frac{1}{R_2/R_1} + \cdots \ ,
\end{align}
in the large width limit ($R_2/R_1 \gg 1$).
What we observed in the previous subsection is nothing but this asymptotic form consistent with numerical result shown in Fig.\,\ref{fig:masslessI}.

\section{Holographic entanglement entropy }\label{ss:HEE}
In this section, we examine the entanglement entropy of an annulus in CFT$_3$ and a gapped system holographically described by the Einstein-Hilbert gravity in the (asymptotically) AdS$_4$ space.
The holographic formula \cite{Ryu:2006bv,Ryu:2006ef} 
\begin{align}\label{HEE}
  S_A=\underset{\partial\gamma_A=\Sigma}{\mathrm{min}}\,\frac{\mathrm{Area}[\gamma_A]}{4G_N}\ ,
\end{align}
associates the entropy of a given region $A$ to the area of a codimension-two minimal surface $\gamma_A$ homologous to the region $A$, i.e., $\partial\gamma_A = \Sigma$. 
Fig.\,\ref{fig:disk_phases} illustrates the cases for $A$ being a disk.

If there are multiple extremal surfaces, we always pick one of them with least area according to the formula \eqref{HEE}, which yields a transition between minimal surfaces as we vary a parameter such as a gap scale.
In this sense, each extremal surface can be regarded as a phase in QFT as we will see in the following.

\subsection{The AdS$_4$ background}
We start with CFT$_3$ dual to the AdS${}_4$ background
\begin{align}
  ds^2=L^2\frac{dz^2-dt^2+dr^2+r^2d\theta^2}{z^2}\ ,
\end{align}
with the AdS radius $L$. 
The original CFT$_3$ is interpreted to live on the boundary $z=0$ (or at $z=\epsilon \ll 1$ if UV regularization is needed).

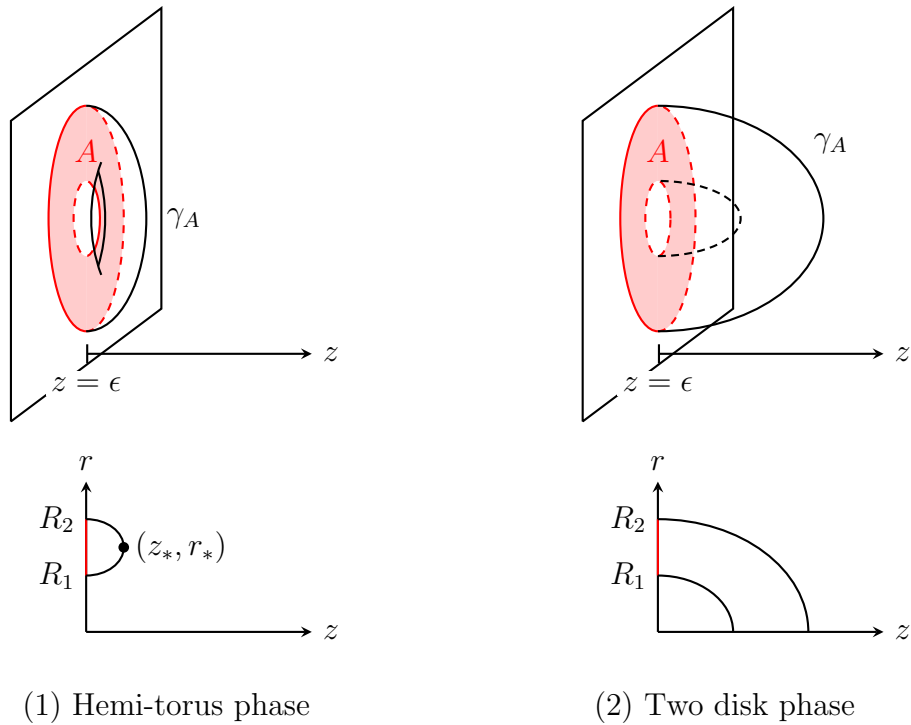
\begin{figure}[htbp]
\centering
\begin{tikzpicture}[thick,>=stealth]
   \tikzstyle{ann} = [fill=white,inner sep=2pt]
   \draw (-1,-1.2)--(1,0.3)--(1,4.3)--(-1,2.8)--(-1,-1.2);
   \draw[dashed,red,fill=red!20] (0,0) arc (-90:90:0.5 and 1.5);
   \draw[red,fill=red!20] (0,0) arc (270:90:0.5 and 1.5);
   \draw[dashed,red,fill=white] (0,1.5) ellipse (0.167 and 0.5);
   \draw[red,fill=white] (-0.15,1.5) ++(-60:0.5) arc (-60:60:0.167 and 0.5);
   \draw[] (-1.2,1.5) ++(25:1.5) arc (25:-25:1 and 1.5);
   \draw[] (1.5,1.5) ++(210:1.5) arc (210:150:1 and 1.5);
   \draw[] (0,0) arc (-90:90:0.8 and 1.5);
   \draw (0,2.4) node {$\textcolor{red} A$};
   \draw (1.3,1.5) node {$\gamma_A$};
   \draw[|->] (0,-0.3) -- (3,-0.3) node[right] {$z$};
   \draw (0,-0.7) node[ann] {$z=\epsilon$};

   \draw[->] (0,-4) -- (0,-2) node[above] {$r$};
   \draw[->] (0,-4) -- (3,-4) node[right] {$z$};
   \draw[red] (0,0.75-4) -- (0,1.5-4);
   \draw[] (0,0.75-4) node[left] {$R_1$} arc (-90:90:0.5 and 0.75/2) node[left] {$R_2$};
   \draw[] (0.5,0.75-4+0.75/2) node[fill,circle,inner sep=0.05 cm] {};
   \draw[] (0.5,0.75-4+0.75/2) node[right] {$(z_*,r_*)$};   
   \draw (-1,-2-3) node[right] {(1) Hemi-torus phase};
\end{tikzpicture}
\hspace{80pt}
\begin{tikzpicture}[thick,>=stealth]
   \tikzstyle{ann} = [fill=white,inner sep=2pt]
   \draw (-1,-1.2)--(1,0.3)--(1,4.3)--(-1,2.8)--(-1,-1.2);
   \draw[dashed,red,fill=red!20] (0,0) arc (-90:90:0.5 and 1.5);
   \draw[red,fill=red!20] (0,0) arc (270:90:0.5 and 1.5);
   \draw[dashed,red,fill=white] (0,1.5) ellipse (0.167 and 0.5);
   \draw[] (0,0) arc (-90:90:2.2 and 1.5);
   \draw[densely dashed] (0,1) arc (-90:90:1.1 and 0.5);
   \draw (0,2.4) node {$\textcolor{red} A$};
   \draw (2.3,2.5) node {$\gamma_A$};
   \draw[|->] (0,-0.3) -- (3,-0.3) node[right] {$z$};
   \draw (0,-0.7) node[ann] {$z=\epsilon$};

   \draw[->] (0,-4) -- (0,-2) node[above] {$r$};
   \draw[->] (0,-4) -- (3,-4) node[right] {$z$};
   \draw[red] (0,0.75-4) -- (0,1.5-4);
   \draw[] (0,-4) ++(0:2)  arc (0:90:2 and 1.5) node[left] {$R_2$};
   \draw[] (0,-4) ++(0:1) arc (0:90:1 and 0.75) node[left] {$R_1$};

   \draw (-1,-2-3) node[right] {(2) Two disk phase};
\end{tikzpicture}
\caption{Two phases for the minimal surface in the AdS$_4$ background:
connected hemi-torus phase (1) and disconnected two disk phase (2). Here the time $t$ direction is suppressed.
}
\label{fig:CFT_phases}
\end{figure}

The extremal surface respecting the rotational symmetry of the annulus is a solution to the equation of motion for the action
\begin{align}\label{AnnulusHEE}
  I[r(z)]=\frac{\pi L^2}{2G_N}\int dz\, \frac{r(z)\sqrt{1+r'(z)^2}}{z^2} \ ,
\end{align}
with the boundary conditions $r(0)=R_i$ ($i=1,2$) on its ends.
There are two possible extremal surfaces depending on their topologies:
\begin{itemize}
\item 
{\bf Two disk phase} (Fig.\,\ref{fig:CFT_phases} (2)): $\gamma_A$ is the superposition of disconnected two disks, each of them being given by
\begin{align}
r(z)=\sqrt{R_i^2-z^2} \ , \qquad  (i=1,2) \ ,
\end{align}
respectively.
This solution always exists independent of the size of the annulus. 
\item
{\bf Hemi-torus phase} (Fig.\,\ref{fig:CFT_phases} (1)): $\gamma_A$ is a connected extremal surface.
The analytic solution is obtained in the following way \cite{Drukker:2005cu,Dekel:2013kwa,Fonda:2014cca}.
It consists of two branches in the $(r, z)$-plane as
\begin{align}
r =     \begin{cases}
        R_1 \exp\left[ -f_- (z/r) \right] \ , \\
        R_2  \exp\left[ -f_+ (z/r) \right] \ , 
        \end{cases}
\end{align}
where the functions $f_\pm(x)$ are defined using the incomplete elliptic integrals\footnote{The definitions of the incomplete elliptic integrals used here are
\begin{align}
\begin{aligned}
        \BF (x|m) &\equiv \int_0^x d\theta \frac{1}{\sqrt{1-m\sin^2 \theta}} \ , \\
        \Pi (n, x | m) &\equiv \int_0^x d\theta \frac{1}{(1-n\sin^2\theta) \sqrt{1-m\sin^2\theta}} \ ,
\end{aligned}
\end{align}
and $\BK(m) \equiv \BF (\pi/2|m)$ and $\Pi (n|m)\equiv \Pi (n,\pi/2| m)$.
}
by
\begin{align}
f_\pm (x) = \frac{1}{2}\log(1+x^2) \pm \eta\, x_m \left[ \BF\left( \omega(x)|\eta^2\right) - \Pi \left( 1 -\eta^2,\omega(x) | \eta^2\right)\right] \ ,
\end{align}
with the range $0\le x \le x_m \equiv \sqrt{\frac{2\eta^2-1}{1-\eta^2}}$ and $\omega(x) = \arcsin \left[ \frac{x/x_m}{\sqrt{1 - \eta^2 (1-x/x_m )}} \right]$.
The parameter $\eta$ in the range $\eta \in [1/\sqrt{2}, 1]$ is related to the ratio $R_2/R_1$ of the inner and outer radii of the annulus as
\begin{align}\label{Ratio_kappa}
\log (R_2/R_1) = 2\eta \sqrt{\frac{2\eta^2-1}{1-\eta^2}} \left[ \BK(\eta^2) - \Pi (1-\eta^2|\eta^2)\right] \ .
\end{align}
This solution is available only for $(1 \le )R_2/R_1 < 2.724$.
\end{itemize}

\begin{figure}[htbp]
\begin{center}
\includegraphics[width=8cm]{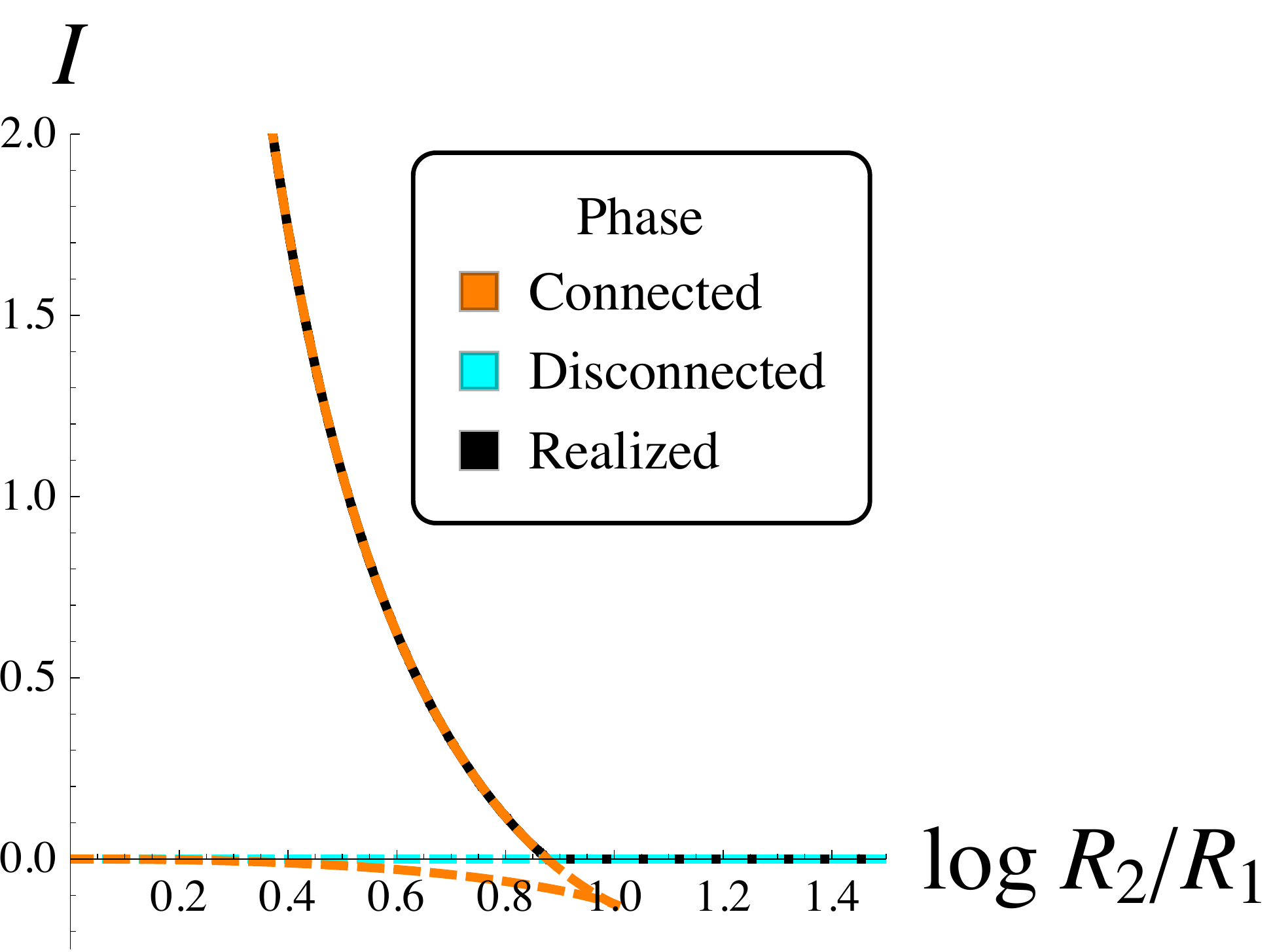}
\end{center}
\caption{The holographic mutual information $I$ across the annulus of radii $R_1$ and $R_2$ for CFT. This mutual information $I=I(R_2/R_1)$ has a phase transition at $(R_2/R_1)_\text{critical}\simeq2.4$, and vanishes for $R_2/R_1>(R_2/R_1)_\text{critical}$ because the disconnected two disk phase is realized.}
\label{fig:confIdual}
\end{figure}

The two disk phase is realized for the large width $R_2/R_1 > 2.724$ where it is the unique solution, while it compete with the hemi-torus phase when $R_2/R_1 < 2.724$.
In order to fix the location of the phase transition, we calculate the mutual information $I$ across the annulus defined by \eqref{MI}.
It is clear in the holographic setup that $I>0$ signifies the hemi-torus phase  
because $I=0$ in the two disk phase.\footnote{The mutual information can vanish only in the large-$N$ limit and there are $O(1/N)$ corrections \cite{Faulkner:2013ana,Engelhardt:2014gca} for finite $N$.
More generally, the mutual information is bounded from below.} 
We benefit from the relevant result of \cite{Fonda:2014cca} to get the mutual information in the hemi-torus phase\footnote{The elliptic integral of the second kind is defined by
\begin{align}
        \BE (m) \equiv \int_0^{\pi/2} d\theta\, \sqrt{1-m \sin^2\theta} \ .
\end{align}
}
\begin{align}\label{MI_HemiTorus}
I_\text{hemi-torus} = \frac{\pi L^2}{G_N} \left[  \frac{\BE (\eta^2) - (1-\eta^2) \BK (\eta^2)}{\sqrt{2\eta^2 -1} - 1} -1\right] \ ,
\end{align}
whose plot is displayed in orange color in Fig.\,\ref{fig:confIdual}. 
It is a two-valued function whose lower branch is always negative and 
the upper branch intersects with $I=0$ at $R_2/R_1 = (R_2/R_1)_\text{critical} \approx 2.4$.
Since the holographic formula \eqref{HEE} selects the non-negative $I$, the physical mutual information is given by $I_\text{hemi-torus}$ for $R_2/R_1 < (R_2/R_1)_\text{critical}$ and $I=0$ for $(R_2/R_1)_\text{critical} < R_2/R_1$.
It has a kink at $R_2/R_1=(R_2/R_1)_\mathrm{critical}$ caused by the phase transition of the extremal surface $\gamma_A$.
Comparing with the general form \eqref{I_CFT} of the mutual information in CFT, Fig.\,\ref{fig:confIdual} demonstrates the monotonicity and convexity \eqref{SSA} of the function $f$ with respect to $\rho=\log(R_2/R_1)$.
In other words, the holographic entanglement entropy of an annulus satisfies the strong subadditivity as guaranteed by the holographic proof based on the minimality of the surfaces \cite{Headrick:2007km}.

\subsection{The CGLP background}
We move onto a gapped theory described by an asymptotically AdS geometry whose IR region (away from the boundary) is capped off.
As a concrete example, we use the CGLP background \cite{Cvetic:2000db} in M-theory dual to a $(2+1)$-dimensional QFT with a gap scale.

The CGLP background is a $(3+1)$-dimensional geometry times a seven-dimensional internal manifold, which asymptotes to the AdS$_4$ space times the Stiefel manifold $V_{5,2}$.
In the Einstein frame, the metric is given by
\begin{align}\label{CGLPmetric}
ds^2 = \alpha (u)  \left[ du^2 + \beta (u) \left(-dt^2 + dr^2 + r^2 d\theta^2\right) \right]  + g_{ij} dy^i dy^j \ , 
\end{align}
where $u$ is the holographic coordinate of the AdS$_4$ ranging from the IR capped-off point $0$ to the UV fixed point $\infty$.
$y^i \,(i=1,\cdots, 7)$ are the coordinates of the internal manifold with a volume
\begin{align}
V(u) = \int \prod_{i=1}^7 dy^i \sqrt{\det g} \ ,
\end{align}
vanishing at $u=0$. 
The functions in the metric are given by
\begin{align}
\begin{aligned}
        \alpha (u) &= \frac{H(u)^{1/3}c^2(u)}{4}\ , \qquad \beta(u) = \frac{4}{H (u) c^2 (u)}   \ , \\
        V(u) &= \frac{3^{17/8} \pi^4 \varepsilon^{21/4}}{2} H^{7/6} (u) (2+\cosh u)^{3/8} \sinh^{3/2}\left( \frac{u}{2}\right) \sinh^{3/2}u \ , \\
        H(u) &= \frac{L^6}{\varepsilon^{9/2}}2^{3/2} 3^{11/4} \int_{(2+\cosh u)^{1/4}}^\infty \frac{dt}{(t^4 - 1)^{5/2}} \ , \\
         c^2 (u) &= \frac{3^{7/4} \varepsilon^{3/2}\cosh^3 (u/2)}{2(2+\cosh u)^{3/4}} \ ,
\end{aligned}
\end{align}
with two dimensionful parameters $L$ and $\varepsilon$.
The parameter $L$ is the AdS radius near the boundary, determined by the number of M2-branes $N$ and the Planck length $\ell_p$ as
$L\equiv 3^{-2/3}2\pi^{1/3} \ell_p N^{1/6}$.
The parameter $\varepsilon$, defining the size of deformation \cite{Klebanov:2010qs}, has mass dimension $-4/3$, letting $H$ be dimensionless. 
$V$ appears to depend on $\varepsilon$, but does not indeed.
By rescaling the boundary coordinates $(t,r)$ appropriately, one can remove $\varepsilon$ completely from the metric if one wishes.

Let us take a look at the UV behavior of the metric \eqref{CGLPmetric} for a moment.
When $u$ is close to the UV cutoff $u\to\Lambda\gg 1$, the function $H(u)$ becomes
\begin{align}
H(u) \to 2^{15/4} 3^{3/4}\, L^6 \,e^{-9u/4} \ ,
\end{align}
and the other functions approach
\begin{align}
\begin{aligned}
\alpha(u) &\to \frac{9}{16} L^2 \ , &\qquad \beta(u)&\to 2^{3/2}3^{-5/2}\,L^{-6}\,e^{3u/2} \ , \\
V(u)&\to 3^3\pi^4 L^{21/2}\ , &\qquad c^2(u) &\to 2^{-13/4}3^{7/4} e^{3u/4} \ .
\end{aligned}
\end{align}
The transformation $z = 2^{5/4}3^{1/4}L^3 e^{-3u/4}$ takes the metric to the Poincar{\'e} coordinates of the AdS$_4$ space near the boundary
\begin{align}
ds^2  \to L^2 \frac{dz^2 - dt^2 + dr^2 + r^2 d\theta^2}{z^2} + \cdots \ .
\end{align}

\begin{figure}[htbp]
\centering
\begin{tikzpicture}[thick,>=stealth]
   \tikzstyle{ann} = [fill=white,inner sep=2pt]
   \draw (-1,-1.2)--(1,0.3)--(1,4.3)--(-1,2.8)--(-1,-1.2);
   \draw[dashed,red,fill=red!20] (0,0) arc (-90:90:0.5 and 1.5);
   \draw[red,fill=red!20] (0,0) arc (270:90:0.5 and 1.5);
   \draw[dashed,red,fill=white] (0,1.5) ellipse (0.167 and 0.5);
   \draw[domain=0:2] plot (\x,{2.9+0.1*cos(90*\x)});
   \draw[domain=0:2] plot (\x,{0.1-0.1*cos(90*\x)});
   \draw[] (2,1.5) ellipse (0.43 and 1.3);
   \draw[densely dashed] (0,1) arc (-90:90:1.1 and 0.5);
   \draw (0,2.4) node {$\textcolor{red} A$};
   \draw (1.6,3.5) node {$\gamma_A$};
   \draw[|-] (0,-0.3) -- (2,-0.3);
   \draw[|->] (2,-0.3) -- (3,-0.3) node[right] {$z$};
   \draw (0,-0.8) node[ann] {$z=\epsilon$};
   \draw (2.1,-0.8) node[ann] {$z=z_0$};

   \draw[->] (0,-4) -- (0,-2) node[above] {$r$};
   \draw[->] (0,-4) -- (3,-4) node[right] {$z$};
   \draw[dotted] (2,-4) (2,-4) node[below] {$z_0$} -- (2,-2);
   \draw[red] (0,0.75-4) -- (0,1.5-4);
   \draw[domain=2:0] plot (\x,{2.9+0.1*cos(90*\x)-5.5}) node[left] {$R_2$};
   \draw[] (0,-4) ++(0:1) arc (0:90:1 and 0.75) node[left] {$R_1$};

   \draw (-1,-2-3) node[right] {(3) Disk-cylinder phase};
\end{tikzpicture}
\hspace{80pt}
\begin{tikzpicture}[thick,>=stealth]
   \tikzstyle{ann} = [fill=white,inner sep=2pt]
   \draw (-1,-1.2)--(1,0.3)--(1,4.3)--(-1,2.8)--(-1,-1.2);
   \draw[dashed,red,fill=red!20] (0,0) arc (-90:90:0.5 and 1.5);
   \draw[red,fill=red!20] (0,0) arc (270:90:0.5 and 1.5);
   \draw[dashed,red,fill=white] (0,1.5) ellipse (0.167 and 0.5);
   \draw[domain=0:2] plot (\x,{2.9+0.1*cos(90*\x)});
   \draw[domain=0:2] plot (\x,{0.1-0.1*cos(90*\x)});
   \draw[] (2,1.5) ellipse (0.43 and 1.3);
   \draw[densely dashed,domain=0:1.6] plot (\x,{1.9+0.1*cos(90*\x)});
   \draw[domain=1.6:2] plot (\x,{1.9+0.1*cos(90*\x)});
   \draw[densely dashed,domain=0:1.6] plot (\x,{1.1-0.1*cos(90*\x)});
   \draw[domain=1.6:2] plot (\x,{1.1-0.1*cos(90*\x)});
   \draw[] (2,1.5) ellipse (0.1 and 0.3);
   \draw (0,2.4) node {$\textcolor{red} A$};
   \draw (1.6,3.5) node {$\gamma_A$};
   \draw[|-] (0,-0.3) -- (2,-0.3);
   \draw[|->] (2,-0.3) -- (3,-0.3) node[right] {$z$};
   \draw (0,-0.8) node[ann] {$z=\epsilon$};
   \draw (2.1,-0.8) node[ann] {$z=z_0$};
      
   \draw[->] (0,-4) -- (0,-2) node[above] {$r$};
   \draw[->] (0,-4) -- (3,-4) node[right] {$z$};
   \draw[dotted] (2,-4) node[below] {$z_0$} -- (2,-2);
   \draw[red] (0,0.75-4) -- (0,1.5-4);
   \draw[domain=2:0] plot (\x,{2.9+0.1*cos(90*\x)-5.5}) node[left] {$R_2$};
   \draw[domain=2:0] plot (\x,{2.9+0.1*cos(90*\x)-6.25}) node[left] {$R_1$};
   
   \draw (-1,-2-3) node[right] {(4) Two cylinder phase};
\end{tikzpicture}
\caption{Two new disconnected phases for the minimal surface in the CGLP background:
disk-and-cylinder phase (3) and two cylinders phase (4).
In the Poincar\'{e} coordinate $z=2^{5/4}3^{1/4}L^3 e^{-3u/4}$, the  UV boundary $u=\Lambda$ corresponds to $z=\epsilon=2^{5/4}3^{1/4}L^3 e^{-3\Lambda/4}$ and the IR capped-off point $u=0$ corresponds to $z=z_0=2^{5/4}3^{1/4}L^3$.
}
\label{fig:gapped_phases}
\end{figure}
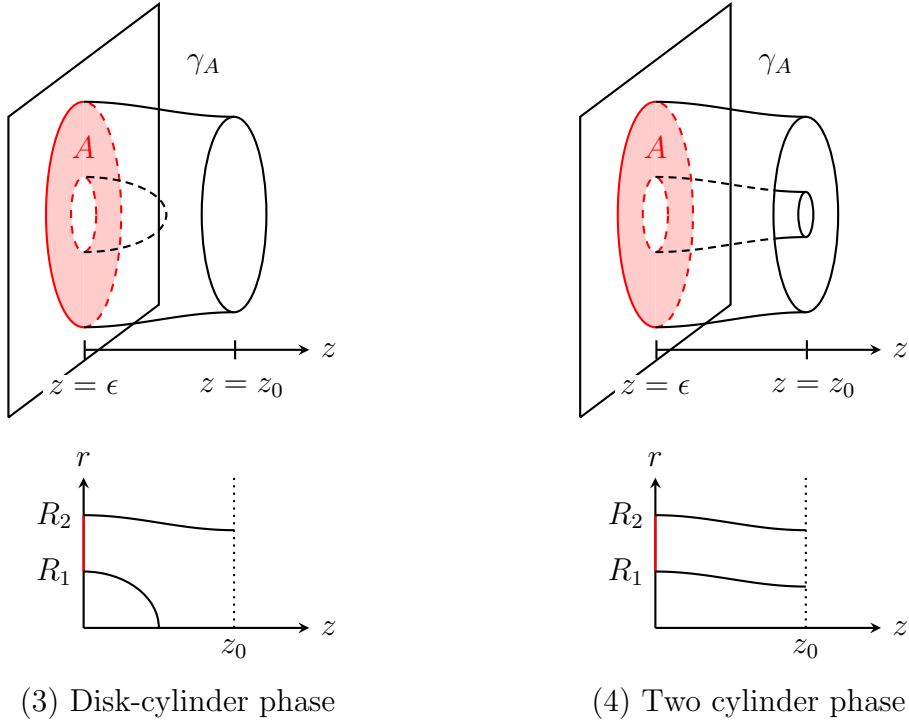

Since the extremal surface for a small annulus localizes near the boundary, the entanglement entropy remains to have the previous two phases shown in Fig.\,\ref{fig:CFT_phases} in the CGLP background.
In addition, there are new phases for a large annulus whose minimal surfaces can reach and terminate on the IR cap-off as we describe below.
\begin{figure}[htbp]
\centering
\begin{tikzpicture}[thick,>=stealth]
   \tikzstyle{ann} = [fill=white,inner sep=2pt]
   \draw (-1,-1.2)--(1,0.3)--(1,4.3)--(-1,2.8)--(-1,-1.2);
   \draw[dashed,red,fill=red!20] (0,0) arc (-90:90:0.5 and 1.5);
   \draw[red,fill=red!20] (0,0) arc (270:90:0.5 and 1.5);
   \draw[] (0,0) arc (-90:90:2.2 and 1.5);
   \draw (0,1.5) node {};
   \draw (2.3,2.5) node {$\gamma_\text{disk}$};
   \draw[|->] (0,-0.3) -- (3,-0.3) node[right] {$z$};
   \draw (0,-0.7) node[ann] {$z=\epsilon$};

   \draw[->] (0,-4) -- (0,-2) node[above] {$r$};
   \draw[->] (0,-4) -- (3,-4) node[right] {$z$};
   \draw[red] (0,-4) -- (0,1.5-4);
   \draw[] (2,-4) node[fill,circle,inner sep=0.05 cm]{};
   \draw[] (0,-4) ++(0:2) node[below]{$z_*$}  arc (0:90:2 and 1.5) node[left] {$R$};

   \draw (-0.5,-2-3) node[right] {($a$) disk phase};
\end{tikzpicture}
\hspace{100pt}
\begin{tikzpicture}[thick,>=stealth]
   \tikzstyle{ann} = [fill=white,inner sep=2pt]
   \draw (-1,-1.2)--(1,0.3)--(1,4.3)--(-1,2.8)--(-1,-1.2);
   \draw[dashed,red,fill=red!20] (0,0) arc (-90:90:0.5 and 1.5);
   \draw[red,fill=red!20] (0,0) arc (270:90:0.5 and 1.5);
   \draw[domain=0:2] plot (\x,{2.9+0.1*cos(90*\x)});
   \draw[domain=0:2] plot (\x,{0.1-0.1*cos(90*\x)});
   \draw[] (2,1.5) ellipse (0.43 and 1.3);
   \draw (0,1.5) node {};
   \draw (1.7,3.2) node {$\gamma_\text{disk}$};
   \draw[|-] (0,-0.3) -- (2,-0.3);
   \draw[|->] (2,-0.3) -- (3,-0.3) node[right] {$z$};
   \draw (0,-0.8) node[ann] {$z=\epsilon$};
   \draw (2.1,-0.8) node[ann] {$z=z_0$};
      
   \draw[->] (0,-4) -- (0,-2) node[above] {$r$};
   \draw[->] (0,-4) -- (3,-4) node[right] {$z$};
   \draw[dotted] (2,-4) node[below] {$z_0$} -- (2,-2);
   \draw[red] (0,-4) -- (0,1.5-4);
   \draw[] (2,-2.7) node[fill,circle,inner sep=0.05 cm]{};
   \draw[domain=0:2] plot (\x,{2.9+0.1*cos(90*\x)-5.5}) node[right] {$r_*$};
   \draw (0,3-5.5) node[left] {$R$};
   
   \draw (-0.5,-2-3) node[right] {($b$) cylinder phase};
\end{tikzpicture}
\caption{Two phases of the extremal surface in calculating holographic entanglement entropy of disks in the CGLP background:
disk phase ($a$) and cylinder phase ($b$).
In the Poincar{\' e} coordinate $z=2^{5/4}3^{1/4}L^3 e^{-3u/4}$, the  UV boundary $u=\Lambda$ corresponds to $z=\epsilon=2^{5/4}3^{1/4}L^3 e^{-3\Lambda/4}$ and the IR capped-off point $u=0$ corresponds to $z=z_0=2^{5/4}3^{1/4}L^3$.
In the cylinder phase, the extremal surface terminates on the IR capped-off point $z = z_0$.
}
\label{fig:disk_phases}
\end{figure}
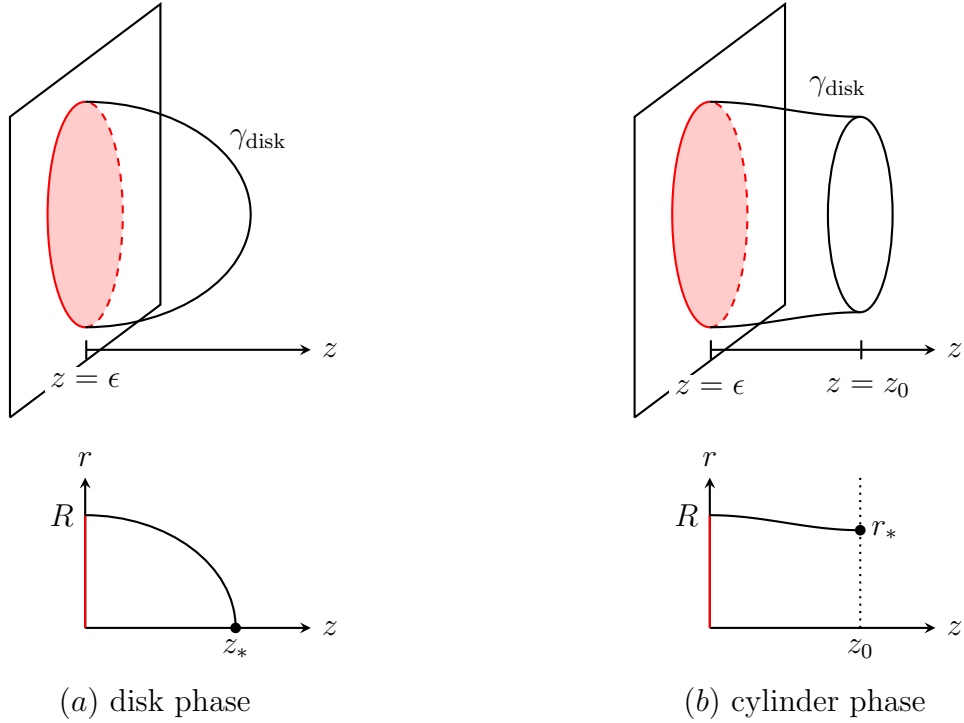
These are superpositions of disk- and cylinder-type solutions for a disk region \cite{Liu:2012eea,Klebanov:2012yf} depicted in Fig.\,\ref{fig:disk_phases}.
They have different topologies as the names suggest, and the cylinder-type solution only exists and dominates for a large radius. 
This resembles the situation for a strip region in a gapped system, which is interpreted as a confinement/deconfinement phase transition \cite{Nishioka:2006gr,Klebanov:2007ws}.
In the present case, the minimal surface switches from the disk-type to the cylinder-type at the critical radius $R = R_\mathrm{critical}\simeq0.72/m$, where $m=\varepsilon^{-3/4}$ is the gap scale determined by the CGLP metric.
Taking into account these facts, we end up with three superposed phases; two disk phase, one disk and one cylinder (disk-cylinder) phase, and two cylinder phase.
The first one has already appeared for CFT in the previous subsection (see Fig.\,\ref{fig:CFT_phases}).
The second and third ones are drawn in Fig.\,\ref{fig:gapped_phases}.
In total, there are the four phases for the annulus in the CGLP background:
\begin{enumerate}
\item[(1)]
the hemi-torus phase (Fig.\,\ref{fig:CFT_phases} (1)) for $R_2-R_1\lesssim 1/m$.
\item[(2)]
the two disk phase (Fig.\,\ref{fig:CFT_phases} (2)) for $R_1,R_2<R_\mathrm{critical}$,
\item[(3)]
the disk-cylinder phase (Fig.\,\ref{fig:gapped_phases} (3)) for $R_1<R_\mathrm{critical}<R_2$,
\item[(4)]
the two cylinder phase (Fig.\,\ref{fig:gapped_phases} (4)) for $R_\mathrm{critical}<R_1,R_2$,
\end{enumerate}

There are apparently overlaps between the first phase and the others, where the one with the least entropy is realized.
To classify the phase structure, we calculate the holographic entanglement entropy in a similar way to the previous pure AdS case.
The rotational symmetry lets us assume the radial coordinate $r$ of the extremal surface $\gamma_A$ as a (two-branched) function $r_\pm=r_\pm(u)$ of the holographic coordinate $u$. 
The area functional becomes
\begin{align}\label{AreaFunctional}
\begin{aligned}
I\left[r(u)\right] = \frac{\pi}{2G_N}\sum_\pm \int du\, r_\pm(u)\, g(u) \sqrt{1 + \beta(u)(r_\pm'(u))^2}  \ ,
\end{aligned}
\end{align}
with $g(u) = V(u) \alpha(u)\beta^{1/2}(u)$.
The extremal surface $r=r(u)$ should satisfy the equation of motion
\begin{align}\label{eom_EE}
2 \, g(u) \sqrt{1+  \beta (u)  (r'(u))^2}= \partial_u \left[\frac{r(u) g(u)\beta(u) r'(u)}{\sqrt{1+  \beta (u) (r'(u))^2}}\right] \ ,
\end{align}
with the boundary conditions $r_+(\infty)=R_2$ and $r_-(\infty)=R_1$.
In contrast to the CFT case, the analytic solution remains to be known. 
Instead, we employ the numerical calculation by the ``shooting method''.

\begin{itemize}
\item
In the hemi-torus phase, we solve the equation of motion \eqref{eom_EE} from the tip $(r,u)=(r_*,u_*)$ where the two branches meet and have an expansion
\begin{align}
r_\pm(u)= r_* \pm 2 \sqrt{\frac{g(u_*)}{g(u_*)\beta' (u_*) +2 g'(u_*)\beta(u_*) }}  \sqrt{u-u_*}  + O((u-u_*)^{3/2})\ .
\end{align}
The radii of the annulus $(R_2,R_1)=(r_+(u=\infty),r_-(u=\infty))$ are functions of $(r_*,u_*)$, respectively.
\item
In the three disconnected phases, the extremal surfaces $\gamma_A$ for the annulus are obtained just by summing the two extremal surfaces $\gamma_{\text{disk}(R_1)}$ and $\gamma_{\text{disk}(R_2)}$ for two disks of radii $R_1$ and $R_2$.
The extremal surface $\gamma_{\text{disk}(R)}$ for a disk in the CGLP metric was obtained \cite{Klebanov:2012yf} as follows.
The disk-type solution can be constructed by solving the equation of motion \eqref{eom_EE} from the tip of the disk $(r,u)=(0,u_*)$, where the extremal surface shrinks as
\begin{align}\label{DiskType}
r(u) = 2 \sqrt{\frac{2g(u_*)}{2\beta (u_*)g'(u_*) + g(u_*) \beta (u_*)} }\sqrt{u-u_*} + O((u-u_*)^{3/2}) \ .
\end{align}
On the other hand,  the cylinder-type solution extends to the IR capped-off point $u=0$ and we solve the equation of motion \eqref{eom_EE} from $(r,u)=(r_*,0)$ where the extremal surface terminates and behaves as
\begin{align}\label{CylinderType}
r(u) = r_* + \frac{1}{8 r_*\beta(0)} u^2 + O(u^{3}) \ .
\end{align}
The disk radius $R=r(\infty)$ is given as a function of $u_*$ or $r_*$, respectively.
\end{itemize}

\begin{figure}[htbp]
\begin{center}
  \includegraphics[width=8cm]{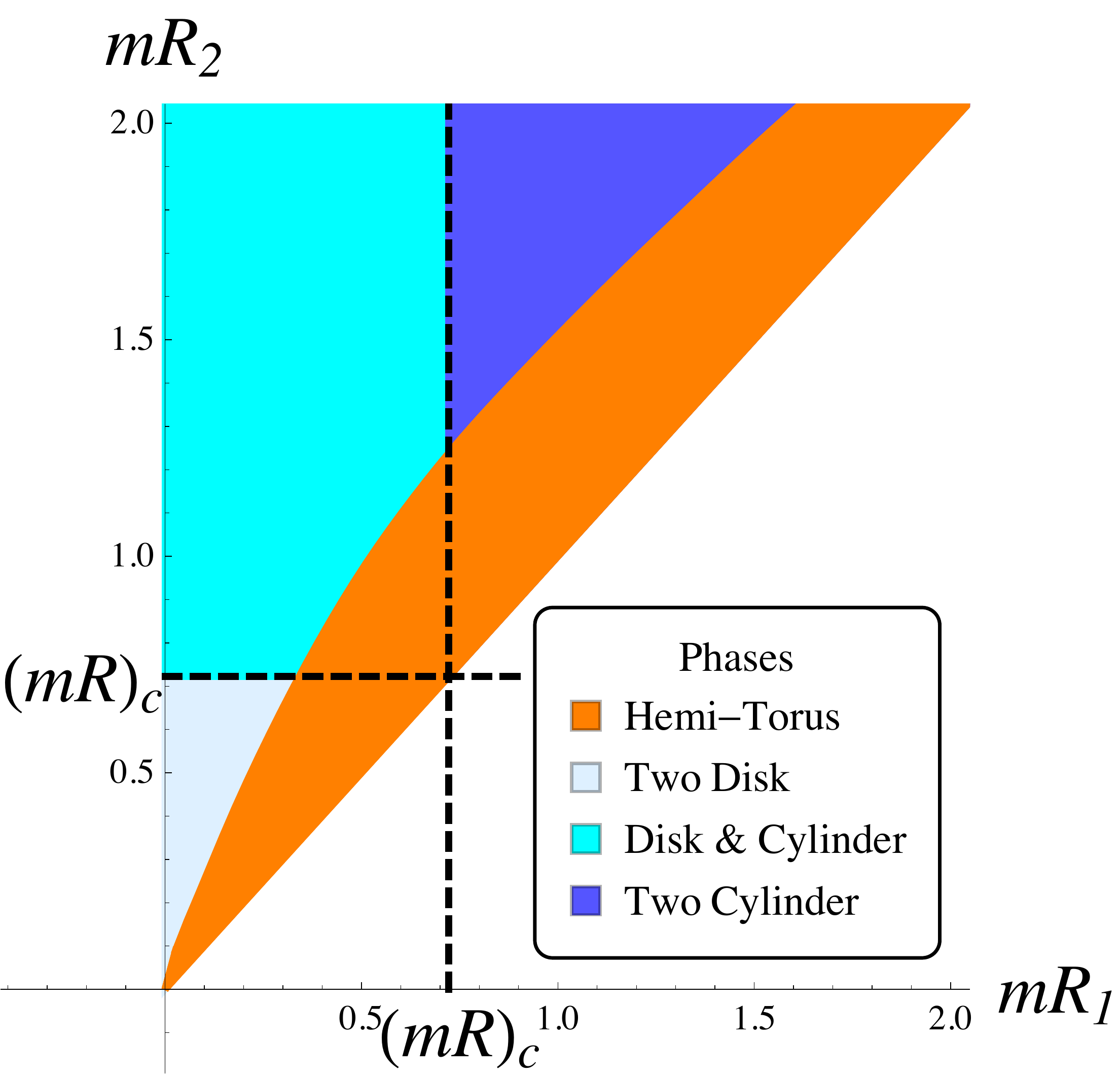}
\end{center}
 \caption{The phase diagram of the entanglement entropy for an annulus of radii $R_1$ and $R_2$. 
The hemi-torus phase is favored when the width of the annulus is small compared to the gap scale.
}
\label{fig:PhaseDiagram}
\end{figure}

After solving the equation of motion numerically, we compare the holographic entanglement entropies \eqref{AreaFunctional} between the four phases.
The resulting phase diagram is presented in Fig.\,\ref{fig:PhaseDiagram}.
It shows that the hemi-torus phase is realized when the width of the annulus is small against the gap scale. 
Note that there is no phase for $R_2/R_1 <1$ since $R_2$ is the outer radius of the annulus.

\begin{figure}[htbp]
\begin{center}
  \includegraphics[width=8cm]{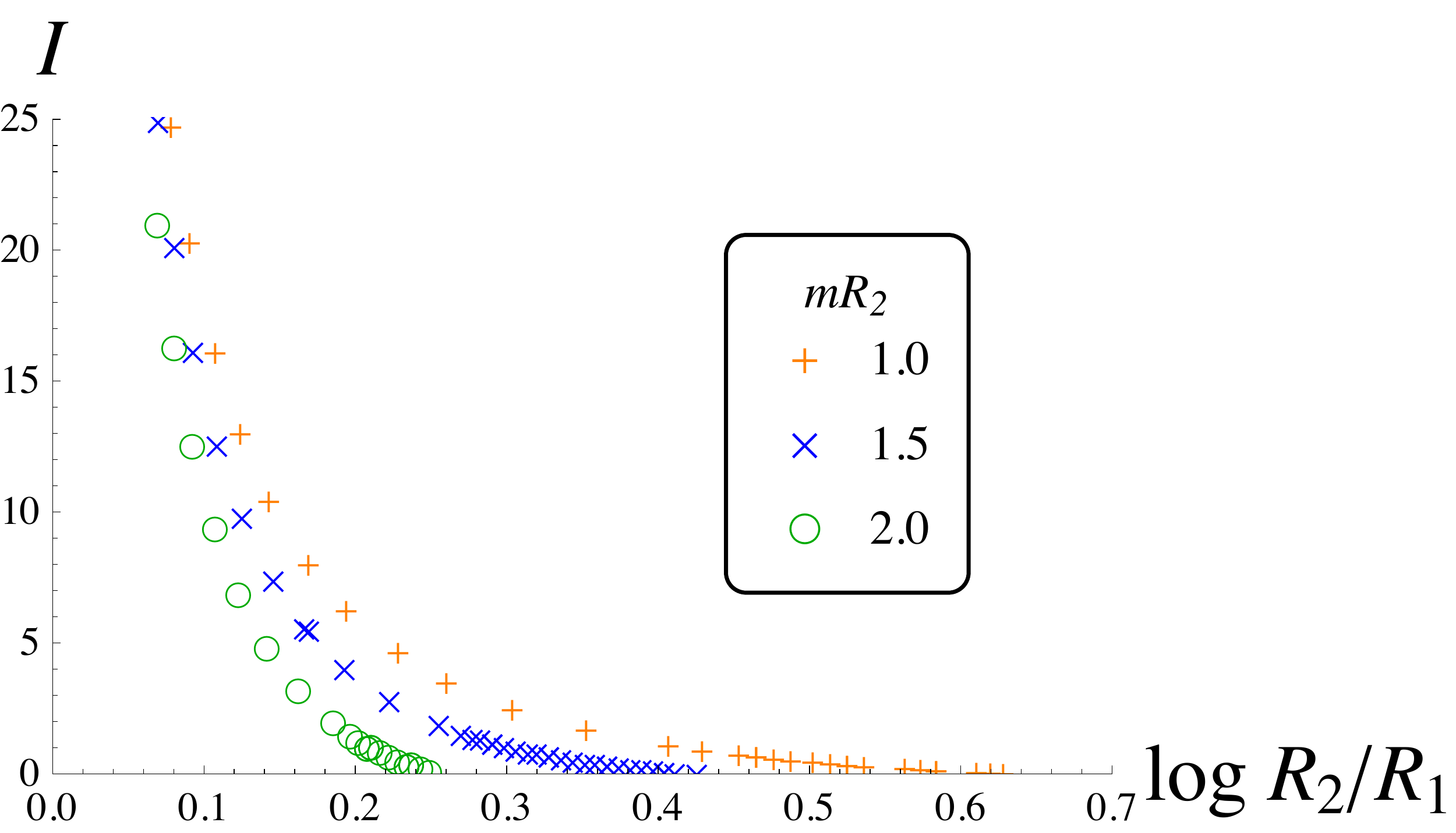}
\end{center}
 \caption{The holographic mutual information $I$ across the annulus of radii $R_1$ and $R_2$ in a gapped theory holographically described by CGLP metric. 
$I=I(mR_1,mR_2)$ vanishes for large $R_2/R_1$ because the phase becomes disconnected, and monotonically decreases with the mass $m$ increased.}
\label{fig:gappedIdual}
\end{figure}

The mutual information \eqref{MI} across the annulus vanishes in all the disconnected phases, and $I>0$ only in the hemi-torus phase. 
Fig.\,\ref{fig:gappedIdual} shows $I$ as a function of $\log (R_2/R_1)$ with $mR_2$ fixed. 
It is positive and decreases as $R_2/R_1$ becomes large, but vanishes at some point due to the phase transition from the hemi-torus phase to a disconnected phase. 
It is also monotonically decreasing with the mass for a fixed $R_2/R_1$.
We will discuss the mass dependence of the mutual information in the next section.

\section{Universal behaviors}\label{ss:Universal}
In the last two sections, we have dealt with the annulus entropies $S_A(R_1,R_2)$ or the mutual informations $I$ across the annulus for the free massive scalar theory and the holographic model. 
In this section, we will compare these two cases, and attempt to identify universal behaviors of entanglement entropy. 

First we consider the small width limit of the mutual information in CFT.
From the field theory result, we anticipate \eqref{ThinMI} holds even in the holographic model. 
Since the hemi-torus phase is always favored in the small width limit, we can make use of the relations \eqref{Ratio_kappa} and \eqref{MI_HemiTorus}.
A short calculation yields the small width behavior \eqref{ThinMI} with the coefficient $\kappa_\text{hol}$ given by\footnote{
See \cite{Agon:2015mja} as a recent related work.
}
\begin{align}
         \kappa_\text{hol} \equiv \frac{L^2 \Gamma[3/4]^4}{2\pi G_N} \ .
\end{align}
It is plausible that $\kappa$ in \eqref{ThinMI} counts the effective degrees of freedom in a given QFT 
because it is proportional to the number of fields in free field theories which characterize the UV fixed point detected by the small width limit of the mutual information.
Indeed, the $\kappa_\text{hol}$ in the holographic model decreases under any RG flow thanks to the holographic $c$-theorem \cite{Girardello:1998pd,Freedman:1999gp,Myers:2010xs,Myers:2010tj} that provides the constraint $L_\text{UV} \ge L_\text{IR}$ for the AdS radii in the UV and IR fixed points.
Similar story may hold for the mutual information through two concentric $(d-2)$-sphere separated by a short distance $\delta$ which behaves as $I\simeq \kappa \,\text{Area}(S^{d-2})/\delta^{d-2}$ in $d\ge 4$ dimensions \cite{Casini:2005zv,Casini:2009sr}.
We do not explore this possibility in this paper, but hope to investigate it in the future.

\begin{figure}[htbp]
\begin{center}
  \includegraphics[width=8cm]{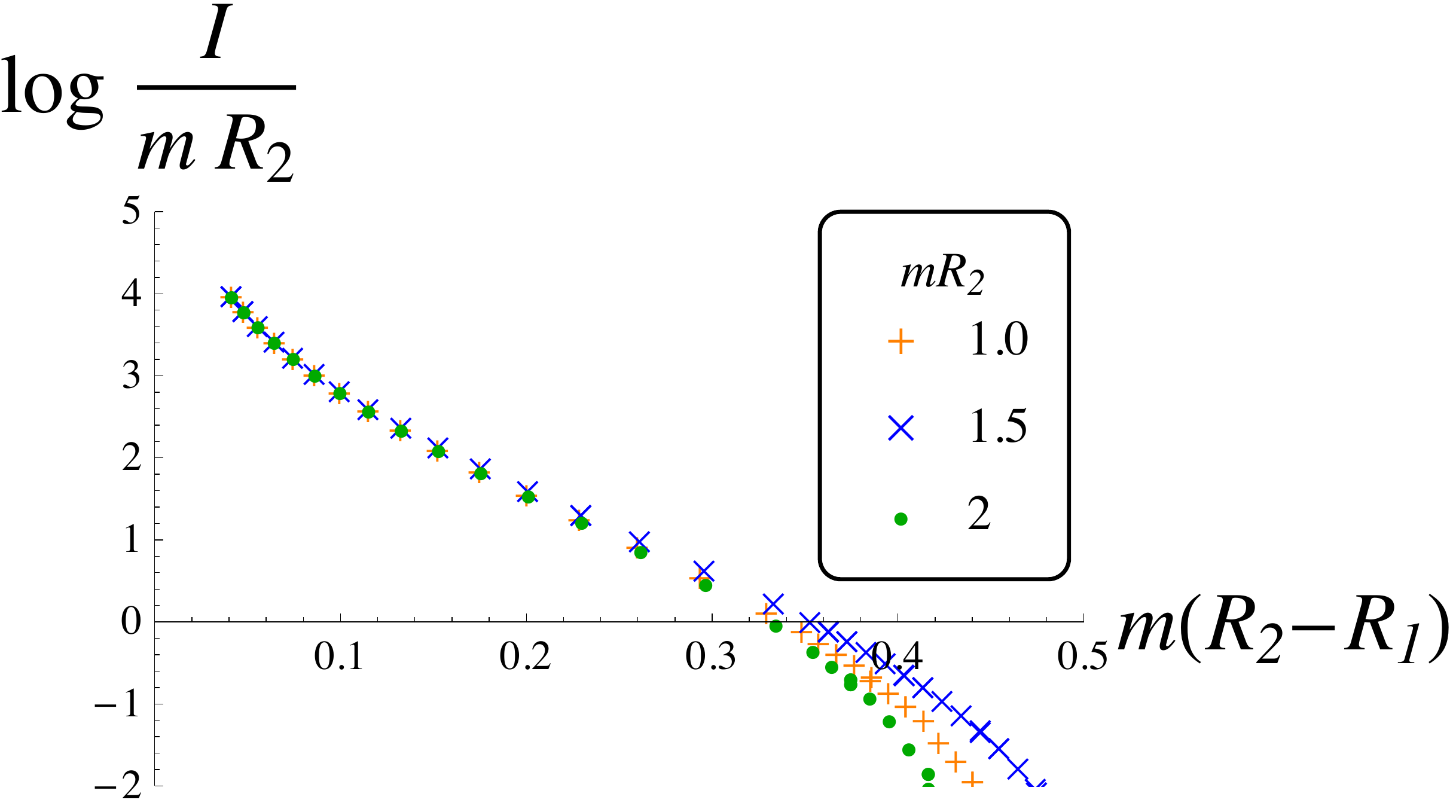}
\end{center}
 \caption{The exponential decay of the holographic mutual information $I$ with the dimensionless width $m(R_2-R_1)$. 
For $0.1 \lesssim m(R_2-R_1)\lesssim0.3$, it shows $I\propto mR_2\exp[-b'\,m(R_2-R_1)]$ with $b'\simeq10$. For $m(R_2-R_1)\gtrsim0.3$, this exponential behavior ends because of the phase transition to the disconnected phases with $I=0$.}
\label{fig:gappedIlogdual}
\end{figure}

In a gapped system, we observed the exponential decay of the mutual information \eqref{I_ExpDecay} for a free massive scalar field. 
It provides a strong evidence for the validity of the ansatz \eqref{S_Gapped} of the entanglement entropy expanded with respect to the inverse of the gap scale, whose coefficients are the integrals of local invariants localized on the entangling surface.
It also implies the existence of an exponentially suppressed correction to the ansatz that will never be seen in the large gap expansion.
It is of interest to see to what extent the ansatz \eqref{S_Gapped} captures the feature of entanglement entropy in a gapped system. 
Actually, our holographic calculation in the CGLP background exhibits the exponential decay of the mutual information as in Fig.\,\ref{fig:gappedIlogdual}.
These observations suggest that the entanglement entropy in a system with a gap $m$ has a power series expansion of $1/m$ with an exponential correction
\begin{align}\label{S_Gapped_Exp}
S_A = \alpha \frac{\ell_\Sigma}{\epsilon} + \beta\, m\,\ell_\Sigma - \gamma_\Sigma + \sum_{n=0}^\infty\frac{c_{2n+1}^\Sigma}{m^{2n+1}} + O(\exp [- m \delta])\ ,
\end{align}
where $\delta$ is proportional to the shortest distance between disjoint entangling surfaces.
This is equivalent to the speculation \eqref{MI_Speculation} for the mutual information of the annulus where $\delta \propto R_2 - R_1$.
We conjecture that \eqref{S_Gapped_Exp} is a universal property in any gapped system.
This resembles the universal thermal corrections in entanglement entropy \cite{Herzog:2012bw,Herzog:2013py,Cardy:2014jwa,Herzog:2014fra,Herzog:2014tfa} and it would be intriguing to find a relationship between them.

\acknowledgments

We are grateful to M.\,Nozaki, K.\,Ohmori, N.\,Shiba, Y.\,Tachikawa and T.\,Takayanagi for valuable discussions.
The work of Y.N. was supported in part by JSPS Research Fellowship for Young Scientists and World Premier International Research Center Initiative (WPI) from the MEXT of Japan.
\appendix

\section{Details of numerical calculations}\label{ss:Numerics}
In this appendix, we summarize the numerical algorithm for calculating the entanglement entropy of the annulus for a free massive scalar field whose action is given by \eqref{ScalarAction}.

\subsection{Radial lattice discretization}
We use the polar coordinates to put the theory on the radial lattice
\begin{align}
ds^2 = - dt^2 + dr^2 + r^2 d\theta^2 \ .
\end{align}
The radial coordinate $r$ is discretized to $N$ points with lattice spacing $a$. 
After the Fourier decomposition along the angular coordinate $\theta$, the lattice Hamiltonian becomes
\begin{align}
H = \frac{1}{2}\sum_{n=-\infty}^\infty\left[ \sum_{i=1}^N \pi^2_{n,i} + \sum_{i,j=1}^N \phi_{n,i} K_n^{i,j} \phi_{n,j} \right] \ ,
\end{align}
where $\phi_{n,i}$ and $\pi_{n,i}$ are the discretized scalar field with angular momentum $n$ on the $i$-th site and its conjugate, respectively.
The matrices $K_n^{i,j}$ depend on the angular momentum and the mass $m$
\begin{align}
K_n^{1,1} = \frac{3}{2} + n^2 + (ma)^2 \ , \qquad K_n^{i,i} = 2 + \frac{n^2}{i^2} + (ma)^2 \ , \qquad K_n^{i,i+1} = K_n^{i+1,i} = - \frac{i+1/2}{\sqrt{i(i+1)}} \ .
\end{align}
These are related to the two-point functions of the scalar fields $(X_n)_{ij}=\langle \phi_{n,i}\phi_{n,j}\rangle$ and the momenta $(P_n)_{ij} = \langle \pi_{n,i}\pi_{n,j}\rangle$ as $X_n = \frac{1}{2}K_n^{-1/2}$ and $P_n = \frac{1}{2}K_n^{1/2}$.

The outer and inner radii of the annulus are chosen to be half-integers in units of the lattice spacing, $R_1/a = r_1 + 1/2$ and $R_2/a = r_2 + 1/2$ with integers $r_1,r_2$.
This choice corresponds to the free boundary condition in the continuum limit.
In our calculation, we vary $r_2$ from 100 to 120 and $r_1$ from $5$ to $r_2-5$.
The entanglement entropy of the annulus $S(R_1,R_2)$ is obtained by using $(r_2 - r_1) \times (r_2 - r_1)$ submatrices $(X_n^{r_1,r_2})_{ij}$ and $(P_n^{r_1,r_2})_{ij}$ of the correlation functions $X_n,P_n$ with the ranges $r_1 + 1 \le i,j \le r_2$ as
\begin{align}\label{Entropy}
S(R_1,R_2) = S_0 + 2\sum_{n=1}^\infty S_n \ ,
\end{align}
where $S_n$ is the contribution from the $n$-th angular mode
\begin{align}\label{Sn}
S_n = \text{tr} \left[ (C_n + 1/2) \log (C_n + 1/2) - (C_n - 1/2) \log (C_n - 1/2) \right] \ ,
\end{align}
with $C_n \equiv \sqrt{X_n^{r_1,r_2} P_n^{r_1,r_2}}$. 
In the following, we describe how to perform this infinite summation
over $n$ under controlled numerical errors.

\subsection{Finite lattice size effect}
To avoid the finite lattice size effect, we repeat the calculation of $S_n$ \eqref{Sn} by changing the lattice size $N$ and fit the results $S_n(N)$
with the asymptotic expansion for large $N$
\begin{align}
S_n(N) = S_n(\infty) + \sum_{k=1}^{k_\mathrm{max}} \frac{a_k}{N^k} \ .
\end{align}
We then read off the constant part $S_n(\infty)$ as the value of $S_n$ in the large-$N$ limit. 
Starting from $N=200$, we increase the lattice size by $\Delta N=20$ until the resultant $S_n(\infty)$ stops changing up to error $\delta=10^{-6}$. 
We choose the fitting parameter $k_\mathrm{max}$ so that the maximum lattice size $N$ is as small as possible.
Typically we find $k_\mathrm{max}=3\sim10$. 

The finite lattice size effect dominates only for small angular momenta $n$ with small masses $ma$.
In our calculation, the maximum lattice size reaches $N\sim1000$ for $n\lesssim10$ in the massless case, but $N=200$ is sufficiently large for $n\gtrsim20$ or $ma\gtrsim0.1$. 
The total numerical error in \eqref{Entropy} can be estimated to be $O(20\delta)\lesssim O(10^{-4})$. 

\subsection{Large angular momentum}
In the large angular momentum limit $n\to \infty$, the correlation matrices $X_n$ and $P_n$ approach almost diagonal matrices \cite{Klebanov:2012yf}.
The products of the submatrices $X_n^{r_1,r_2} P_n^{r_1,r_2}$ almost equal to $1/4$ times unit matrix up to order $1/n^8$.
The nontrivial entries are at the upper-left corners
\begin{align}
\begin{aligned}
(X_n^{r_1,r_2} P_n^{r_1,r_2})^{r_1+1,r_1+1} &= \frac{1}{4} + \frac{r_1^2 (r_1 + 1)^2}{16n^4} - \frac{r_1^2(r_1+1)^2 (2r_1 +1)^2(m^2 +2)}{32n^6} + O(1/n^8) \ , \\
(X_n^{r_1,r_2} P_n^{r_1,r_2})^{r_1+1,r_1+2} &= \frac{r_1^{3} (r_1 + 1)^{3/2} (r_1 +2)^{3/2}}{64n^6} + O(1/n^8) \ , \\
(X_n^{r_1,r_2} P_n^{r_1,r_2})^{r_1+2,r_1+1} &= \frac{r_1^{2}(r_1 +2)^{3/2} (r_1 - 1)^{1/2} (3(r_1+1)^2 -1)}{64n^6} + O(1/n^8) \ .
\end{aligned}
\end{align}
and at the lower-right corners
\begin{align}
\begin{aligned}
(X_n^{r_1,r_2} P_n^{r_1,r_2})^{r_2,r_2} &= \frac{1}{4} + \frac{r_2^2 (r_2 + 1)^2}{16n^4} - \frac{r_2^2(r_2+1)^2 (2r_2 +1)^2 (m^2 +2)}{32n^6} + O(1/n^8) \ , \\
(X_n^{r_1,r_2} P_n^{r_1,r_2})^{r_2,r_2-1} &= \frac{r_2^{3/2} (r_2 - 1)^{3/2} (r_2 + 1)^3}{64n^6} + O(1/n^8) \ , \\
(X_n^{r_1,r_2} P_n^{r_1,r_2})^{r_2-1,r_2} &= \frac{r_2^{1/2}(r_2 -1)^{3/2} (r_2 + 1)^2 (3r_2^2 -1)}{64n^6} + O(1/n^8) \ .
\end{aligned}
\end{align}
Here we restrict the ranges of $r_1,r_2$ to $3\le r_1$ and $r_1+3 < r_2$ to avoid the overlap between the upper-left and lower-right corners, which is satisfied in our set up with $5\le r_1\le r_2-5$. 

The $r_2 - r_1-2$ eigenvalues of the matrix $\sqrt{X_n^{r_1,r_2} P_n^{r_1,r_2}}$ are $1/2+O(1/n^8)$ and the other two are given by
\begin{align}
\frac{1}{2} + c_n^{(a)} - c_n^{(a)} \frac{(2r_a+1)^2 (m^2 +2)}{2n^2} \ , \qquad c_n^{(a)} \equiv \frac{r_a^2 (r_a+1)^2}{16n^4} \ ,\qquad a = 1,2 \ .
\end{align}
Therefore, most of the eigenvalues do not contribute to the $n$-th entanglement entropy \eqref{Sn} up to order $1/n^8$ and 
we obtain
\begin{align}\label{LargeSn}
S_n &= \sum_{a=1,2} \left[ c_n^{(a)} (1-\log c_n^{(a)} ) + \frac{(2r_a +1)^2 (m^2 +2) }{2n^2} c_n^{(a)} \log c_n^{(a)} \right]+ O(1/n^8) \ .
\end{align}
This asymptotic formula is much faster than the direct calculation of \eqref{Sn}.

We perform the matrix trace calculation \eqref{Sn} for $n$ less than some large angular momentum $n_*$, and use this asymptotic formula \eqref{LargeSn} for $n\ge n_*$ as long as $S_n(=O(\log n/n^4))$ is larger than the machine precision.
The other higher modes are ignored.

Our $n_*$ is determined as follows.
Let the error of $O(1/n^8)$ in \eqref{LargeSn} be $\mu/n^8$ with $\mu=\mu(m,r_1,r_2)$.
Then the total numerical error in \eqref{Entropy} is estimated to be $\sum_{n_*}^\infty (\mu/n^8)\sim\mu/(7n_*^7)$.
We take $n_*$ to be the angular momentum where the asymptotic formula \eqref{LargeSn} agrees with the matrix trace calculation \eqref{Sn} up $7\delta/n$. 
Then $\mu/n_*^8\lesssim7\delta/n_*$ holds and the total numerical error in \eqref{Entropy} is bounded by $\sum_{n_*}^\infty (\mu/n^8)\sim\mu/(7n_*^7)\lesssim\delta$.
In this way, we can handle the numerical error within $O(\delta)$.


\bibliographystyle{JHEP}
\bibliography{EE_Annulus}

\end{document}